\documentclass[12pt, draftclsnofoot,onecolumn]{IEEEtran}
\usepackage{amsmath, amssymb, cite, tikz, graphicx, xspace, mathrsfs, mathtools, psfrag, chemarrow, subfigure,color,soul,graphicx,mathtools}
\usepackage{kbordermatrix}
\usepackage{enumitem}
\usepackage{bbm}
\usepackage[utf8]{inputenc}
\usepackage{booktabs}
\usepackage{amscd}
\usepackage{amsmath}
\usepackage{amssymb}
\usepackage{amsthm}

\usepackage{epsfig}
\usepackage{verbatim}
\usepackage{graphicx}
\usepackage{amsthm}
\usepackage{color}
\usepackage{ multirow }
\usepackage[noend]{algpseudocode}
\usetikzlibrary{patterns}
\usetikzlibrary{angles} 
\usepackage{eucal}
\usepackage{algorithm}
\usepackage{tikz}
\usetikzlibrary{shapes, arrows, decorations.markings, arrows.meta}
\usetikzlibrary{patterns} 
\usetikzlibrary{arrows,automata,positioning,chains,calc}
\usetikzlibrary{quotes,angles}
\newtheorem*{remark}{Remark}
\makeatletter
\def\BState{\State\hskip-\ALG@thistlm}
\makeatother
\newtheorem{theorem}{Theorem}
\newtheorem{lemma}[theorem]{Lemma}
\begin{document}
\setlength{\abovecaptionskip}{-3pt}
\setlength{\belowcaptionskip}{1pt}
\setlength{\floatsep}{1ex}
\setlength{\textfloatsep}{1ex}
\title{Real-time Reconstruction of Markov Sources and Remote Actuation over Wireless Channels}
\author{Mehrdad Salimnejad, Marios Kountouris, \IEEEmembership{Fellow, IEEE}, and Nikolaos Pappas, 
\IEEEmembership{Senior Member, IEEE}.
\thanks{M. Salimnejad and N. Pappas are with the Department of Computer and Information Science Linköping University, Sweden, email: \{\texttt{mehrdad.salimnejad, nikolaos.pappas\}@liu.se}. M. Kountouris is with the Communication Systems Department at EURECOM, Sophia-Antipolis, France, email: \texttt{marios.kountouris@eurecom.fr}. \\ The work of M. Salimnejad and N. Pappas has been supported in part by the Swedish Research Council (VR), ELLIIT, Zenith, and the European Union (ETHER, 101096526). The work of M. Kountouris has received funding from the European Research Council (ERC) under the European Union’s Horizon 2020 research and innovation programme (Grant agreement No. 101003431).}}

\maketitle
\begin{abstract}
\par In this work, we study the real-time tracking and reconstruction of an information source with the purpose of actuation. A device monitors the state of the information source and transmits status updates to a receiver over a wireless erasure channel. We consider two models for the source, namely an $N$-state Markov chain and an $N$-state Birth-Death Markov process. We investigate several joint sampling and transmission policies, including a semantics-aware one, and we study their performance with respect to a set of metrics. Specifically, we investigate the real-time reconstruction error and its variance, the cost of actuation error, the consecutive error, and the cost of memory error. These metrics capture different characteristics of the system performance, such as the impact of erroneous actions and the timing of errors. In addition, we propose a randomized stationary sampling and transmission policy and we derive closed-form expressions for the aforementioned metrics. We then formulate two optimization problems. The first optimization problem aims to minimize the time-averaged reconstruction error subject to time-averaged sampling cost constraint. Then, we compare the optimal randomized stationary policy with uniform, change-aware, and semantics-aware sampling policies. Our results show that in the scenario of constrained sampling generation, the optimal randomized stationary policy outperforms all other sampling policies when the source is rapidly evolving. Otherwise, the semantics-aware policy performs the best. The objective of the second optimization problem is to obtain an optimal sampling policy that minimizes the average consecutive error with a constraint on the time-averaged sampling cost. Based on this, we propose a \emph{wait-then-generate} sampling policy which is simple to implement.
\end{abstract}

\section{Introduction}
\par Real-time autonomous and cyber-physical systems have recently received significant attention due to their applications in various time-critical scenarios, including swarm robotics, healthcare monitoring systems, autonomous transportation, and environmental monitoring using sensor networks, to name a few. In these systems, a device (e.g., smart sensor) sends status updates to a remote monitor, which tracks the state of the monitoring process to enable timely decision-making or actuation. Therefore, a very relevant yet highly challenging problem in these systems, is to devise operation policies by taking into account the dynamic of the information source, that will enable real-time remote tracking with the purpose of actuation. This is a relevant use case in goal-oriented semantics-aware communications. Most of the existing works on remote tracking propose sampling/scheduling policies that minimize the estimation error or mean square estimation error without considering the importance of generated and transmitted information nor its impact on the utilization of information. In contrast to these works, we propose a set of semantics-aware metrics that quantify the significance of information and capture various characteristics of the system performance.

\subsection{Related work}
\par Several studies have considered the real-time remote tracking problem. The works \cite{xu2004optimal,shi2011sensor,shi2011time,wu2013can,li2010event, VictorNTNU,wu2012event,weimer2012distributed,meng2012optimal,trimpe2014event,leong2016sensor} consider the problem of 
 scheduling in event-triggered estimation where a sensor observes and sends the state of an information source to the receiver when certain events occur. In \cite{xu2004optimal,shi2011sensor,shi2011time,wu2013can,li2010event, VictorNTNU}, event-based triggered policies have been studied in wireless sensor networks to minimize the estimation error while keeping certain constraints on the communication rate. \cite{wu2012event} proposes an event-triggered policy for a remote monitoring system to achieve a balance between communication rate and estimation accuracy.   
In \cite{weimer2012distributed}, the authors propose a global event-triggered transmission strategy in a wireless sensor network to minimize the network energy consumption and the number of transmissions. In \cite{meng2012optimal}, the authors have studied a class of second-order stochastic systems and have proposed optimal sampling policies for periodic and event-based sampling. In \cite{trimpe2014event,leong2016sensor}, the problem of scheduling in event-triggered estimation is considered in a real-time multi-sensor multichannel system. Optimal sampling and transmission strategies for a system with noiseless communication channels have been considered in \cite{imer2005optimal,nayyar2013optimal,chakravorty2014optimal}. In \cite{imer2005optimal}, a sequential estimation problem with limited information is studied where an observer makes a sequential observation of a stochastic process and sends the sample over a noiseless communication channel to the receiver. The work \cite{nayyar2013optimal} considers a remote estimation problem for a noiseless communication system with an energy harvesting sensor and a remote estimator. In \cite{chakravorty2014optimal}, an optimal threshold transmission policy is derived for a noiseless communication system where a sensor makes an observation of a first-order Markov process and sends the sample to the receiver. The works \cite{shi2012scheduling,wu2018optimal,chakravorty2015distortion,chakravorty2016fundamental,sun2019sampling,ornee2021sampling,GuoKostina2022,hui2022real,chakravorty2019remote,lipsa2011remote,huang2019retransmit, pezzutto2022transmission}, consider the properties of monitored processes for real-time monitoring. In \cite{shi2012scheduling}, an optimal transmission policy has been proposed for two Gauss-Markov systems where the states are measured by two sensors. The results of this work have been extended in \cite{wu2018optimal}, where the authors consider a multi-sensor multiple processes scenario and propose an optimal transmission strategy to minimize the sum of the average estimation error covariance of each process. Fundamental limits and trade-offs of remote estimation of Markov processes in real-time communication are studied in \cite{chakravorty2015distortion,chakravorty2016fundamental}. Optimal sampling of stochastic processes for minimizing the mean square estimation error is studied in \cite{sun2019sampling,ornee2021sampling, GuoKostina2022}. In \cite{hui2022real}, a joint sampling and quantization policy is presented for monitoring real-time Brownian motion. In \cite{chakravorty2019remote}, the problem of optimal transmission and estimation is investigated for remote estimation over time-varying packet drop channels when the information source is modeled by finite state Markov chains and first-order autoregressive processes. 
The works \cite{lipsa2011remote,huang2019retransmit, pezzutto2022transmission}
consider a linear time-invariant discrete-time (LTI) model for the monitoring dynamic process and estimating the current state of the process using the previously received measurements and the model of the dynamic process. The goal of these works is to devise sampling policies that minimize the estimation error, without taking into account the usefulness of the information source and its impact on the actuation. Metrics that account for the importance and the effectiveness (semantics) of information \cite{ShannonWeaver49, VoI_USSR, QoI} with respect to the goal of data exchange have been introduced among others in \cite{kountouris2021semantics, Qin22arxiv,tolga21SP, PetarProc2022, GunduzJSAC23}. 
The most closely related work is \cite{pappas2021goal}, where new goal-oriented semantic sampling and communication policies are proposed for the problem of real-time tracking and source reconstruction of a two-state Markov process. 

\subsection{Contributions}
In this work, we extend the problem of real-time remote tracking in a time-slotted wireless communication system considered in \cite{pappas2021goal}. Our key contributions are summarized below.
\begin{enumerate}
    \item We employ two models to describe the source evolution as depicted in Fig. \ref{Infor_Source_DTMC_BDMP}. We consider joint sampling and transmission policies and propose a set of metrics for system performance evaluation.
    \item We derive general expressions for the real-time reconstruction error and characterize its time-average and the variance. We consider the cost of actuation error to investigate the significance or the non-commutative effects of an error at the receiver side since different errors may have different impact on the system. 
    \item We introduce two new timing-aware error metrics, namely the consecutive error and the cost of memory error, which capture the effect on the performance when the system remains in an erroneous state for several consecutive time slots. These are highly relevant in real-time and/or mission-critical applications where being in an erroneous state for a time duration may lead to safety issues or could even have catastrophic consequences for the system or its user. The consecutive error metric captures the temporal sequence/evolution of errors, which can cumulatively have a serious impact on the actuation performance. In the cost of memory error, we consider the memory of actuation error when the system is not in a synced state over several consecutive time slots. 
    \item We formulate and solve two constrained optimization problems to minimize the time-averaged reconstruction error and the average of consecutive error, respectively, subject to a sampling cost constraint. In the first optimization problem, we provide a comparison among the different sampling and transmission policies and we also show under which conditions the semantics-aware policy outperforms the rest, as well as when the proposed randomized stationary policy can be beneficial. In the second optimization problem, our objective is to obtain the optimal sampling policy that minimizes the average consecutive error. As a result, a \emph{wait-then-generate} sampling policy is proposed under which the sampler decides when to perform sampling to minimize the average consecutive error and satisfy the sampling cost constraint.
\end{enumerate}

\section{System Model}
\label{system model}
\par
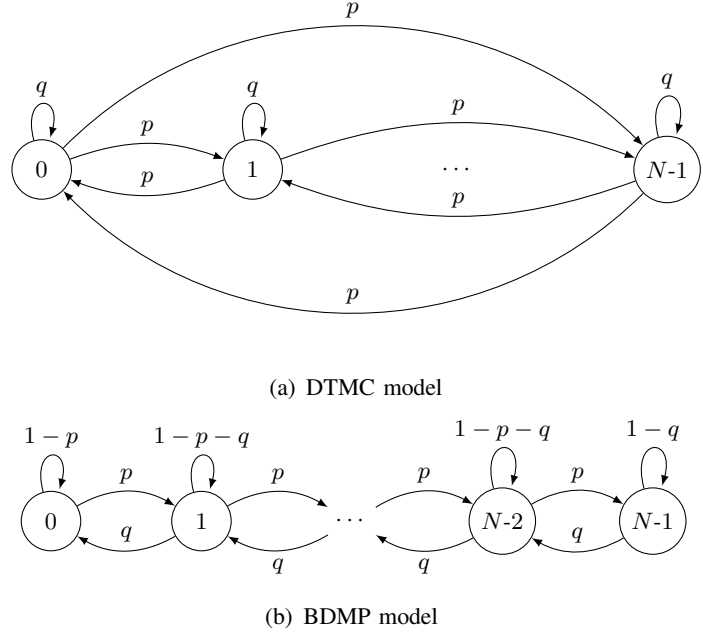
\begin{figure}
	\centering
 \footnotesize
\subfigure[DTMC model]  
{  
         \begin{tikzpicture}[start chain=going left,->,>=latex,node distance=2cm]
         
		\node[state, on chain]                 (N-1){$N$-$1$};
		\node[on chain]                        (g) {$\cdots$};
		\node[state, on chain]                 (2){$1$};
		\node[state, on chain]                 (1){$0$};
		\draw[>=latex]
		
		(1)   edge[loop above] node {$q$} ()
		(1)   edge[bend left=20] node[above] {$p$}(2)
		(1)   edge[bend left=45] node[above]{$p$} (N-1)
		(2)   edge[loop above] node {$q$} ()
		(2)   edge[bend left=20] node[above] {$p$}(1)
		(2)   edge[bend left=20] node[above]{$p$} (N-1)
		(N-1)   edge[loop above] node {$q$} ()
		(N-1)   edge[bend left=20] node[above] {$p$}(2)
		(N-1)   edge[bend left=45] node[above] {$p$}(1)
		;
  
	\end{tikzpicture}
 \label{DTMC_NStates}
}
	\subfigure[BDMP model]{
	\begin{tikzpicture}[start chain=going left,->,>=latex,node distance=2cm,on grid,auto]
		\node[state, on chain]                 (N-1) {$N$-$1$};
		\node[state, on chain]                 (N-2) {$N$-$2$};
		\node[on chain]                        (g) {$\cdots$};
		\node[state, on chain]                 (2) {$1$};
		\node[state, on chain]                 (1) {$0$};
		\draw[>=latex]
		(1)   edge[loop above] node {$1-p$} (1)
		(1) edge  [bend left] node {$p$} (2)
		(2) edge  [loop above] node {$1-p-q$} (2)
		
		(2) edge  [bend left] node[above] {$q$} (1)
		
        (2) edge  [bend left] node {$p$} (g)
		(g) edge  [bend left] node {$q$} (2)
        (g) edge  [bend left] node {$p$} (N-2)
		(N-2) edge  [bend left] node {$q$} (g)
		(N-1) edge  [loop above] node {$1-q$} (N-1)
		(N-2) edge  [loop above] node {$1-p-q$} (N-2)
		(N-2) edge  [bend left] node {$p$} (N-1)
		(N-1) edge  [bend left] node[above] {$q$} (N-2)
		;
	\end{tikzpicture}
 \label{Birth-Death Markov Chain Process}
}
	\vspace*{1ex}
	\caption{The processes describing the evolution of the information source that are considered here.}
 \label{Infor_Source_DTMC_BDMP}
\end{figure}

  We consider a time-slotted communication system in which a sampler performs sampling of a process $X_{t}$ at time slot $t$, and then the transmitter informs the receiver by sending samples over a wireless communication channel. For the information source we consider the following two models. First, an $N$-state discrete time Markov chain (DTMC) depicted in Fig. \ref{DTMC_NStates} in which  the self-transition probability and the probability of transition to another state at time slot $t+1$ are defined as $\text{Pr}\big[X(t+1) = X(t)\big] = q$ and $\text{Pr}\big[X(t+1) \neq X(t)\big] = p$, respectively. Since the process of $X_t$ can have one of $N$ different possible values, we can write $q+(N-1)p = 1$. Second, an $N$-state Birth-Death Markov Process (BDMP) model 
 depicted in Fig. \ref{Birth-Death Markov Chain Process} in which the transition probabilities of $X_{t}$ are given by
\begin{align}
	\mathrm{Pr}\big[X_{t+1} = i\big|X_{t} = i\big]  &=\mathbbm{1}\big(i=0\big)\big(1\!-\!p\big)\!+\mathbbm{1}\big(1\!\leqslant\! i \leqslant\! N\!-\!2\big)\big(1\!-\!p\!-\!q\big)\!+\mathbbm{1}\big(i=N\!-\!1\big)\big(1-q\big),\notag\\
\mathrm{Pr}\big[X_{t+1} = j\big|X_{t} = i\big]  &=\mathbbm{1}\big(0\!\leqslant i\!\leqslant N-2, j=i+1\big)p+\mathbbm{1}\big(1\!\leqslant i\!\leqslant N\!-\!1,j=i-1\big)q.
\end{align}
\par We denote the action of sampling at time slot $t$ by $\alpha^{\text{s}}_{t}$, where $\alpha^{\text{s}}_{t}=1$ if the source is sampled and $\alpha^{\text{s}}_{t} = 0$ otherwise. The action of transmitting a sample is defined as $\alpha^{\text{tx}}_{t}$, where $\alpha^{\text{tx}}_{t}=1$ if a sample is transmitted, otherwise the transmitter remains idle, i.e., $\alpha^{\text{tx}}_{t}=0$.

 \par We assume that the communication channel between transmitter and receiver is subject to small-scale Rayleigh fading and large-scale pathloss attenuation. The received power is given by ${P}_{\text{rx}} = {P}_{\text{tx}}g r^{-\beta}$
where $P_{\text{tx}}$ is the transmitted power, and $g$ is fading power between transmitter and receiver. We assume that $g$ is independent and identically distributed (i.i.d) random variable (RV) with unit mean and cumulative distribution function (CDF) $F_{g}(x) = 1-e^{-x}, x \geqslant 0$. The distance between transmitter and receiver is denoted by $r$, where $\beta > 2$ is the pathloss exponent. At each time slot, the received signal to noise ratio (SNR) is given by $\text{SNR} = \frac{P_{\text{rx}}}{\sigma^2},$
where $\sigma^2$ is the variance of the complex additive white Gaussian noise (AWGN) at the receiver. At time slot $t$, the receiver constructs an estimate of the process $X_{t}$, denoted by $\hat{X}_{t}$. The channel realization $h_{t}$ is equal to $1$ if a sample is successfully decoded by the receiver and $0$ otherwise. A transmitted sample is successfully received if the received SNR exceeds a certain threshold, its probability is given by
\begin{align}
	\label{success_Prob}
	p_{s} = \mathrm{Pr}\big[h_{t}=1\big]= \mathrm{Pr}\big[\text{SNR}>\gamma\big] = \exp\left(-\frac{\gamma\sigma^{2}}{P_{\text{tx}}r^{-\beta}}\right),
\end{align}
where $\gamma$ is the SNR threshold. We consider that a successful transmission is declared to the transmitter using an acknowledgment (ACK) packet. The receiver also sends a negative-ACK packet in the case of a transmission failure. It is assumed that ACK/NACK packets are delivered instantaneously and error free to the transmitter. Therefore, the transmitter has perfect knowledge of the reconstructed source state at time slot $t$, i.e., $\hat{X}_{t}$. We also assume that a sample is discarded when its transmission fails.
\section{Sampling and Transmission Policies and Performance Metrics}
\label{SamplingPoliciesandMetrics}
\par In this section, we introduce four sampling and transmission policies employed in our system, and we define a set of metrics, two newly proposed, to evaluate the performance of the system.

\subsection{Sampling and Transmission Policies}
\label{SamplingPolicies}
\begin{enumerate}
	\item \textbf{Uniform:} sampling is performed periodically every $d$ time slot, independently of the evolution of the source $X_{t}$. Therefore, the sampling time sequences are $\{t_{k} = k d, k\geqslant 1\}$. Although this policy's implementation is simple, several state transitions can be missed between two consecutive sampling events. 
	\item \textbf{Change-Aware:} the generation of a new sample is triggered when a change at the state of the source $X_{t}$ between two consecutive time slots is observed, regardless of whether the system is in synced state or not. This policy does not require an ACK/NACK feedback channel from the receiver to the transmitter.
	\item \textbf{Semantics-Aware:} sample generation is triggered in two cases. First, in the case where at a given time slot $t$ the system is in synced state, i.e., $X_{t}=\hat{X}_{t}$, sampling is performed if a change at the state of the source at time slot $t+1$ occurs, i.e., $X_{t+1}\neq {X}_{t}$. Second, in the case where at time slot $t$, the system is in an erroneous state, i.e., $X_{t}\neq \hat{X}_{t}$, sample acquisition is triggered whenever $X_{t+1}\neq \hat{X}_{t}$. 
	\item \textbf{Randomized Stationary:} sampling is performed in a probabilistic manner at each time slot. As shown later, this simple policy that we propose here, can be optimized under a sampling cost constraint so as to outperform all other sampling policies when the source changes rapidly. 
\end{enumerate}

In Section \ref{OptimizationProblem2}, we propose an additional policy that we define based on the evaluation metrics, thus the discussion is postponed until Section \ref{OptimizationProblem2}.

\subsection{Error Metrics}
\label{metrics}
\begin{enumerate}
	\item \textbf{Real-Time Reconstruction Error:} the real-time reconstruction error is defined as the difference between $X_{t}$ and $\hat{X}_{t}$ at time slot $t$, i.e.,
	\begin{align}
		\label{reconstructed_error}
		E_{t} = \left|X_{t}- \hat{X}_{t}\right|,
	\end{align}
The time-averaged reconstruction error for an observation interval $[1, T]$, with $T$ being a large positive number, is defined as
	\begin{align}
		\label{timeAvg_Error}
\bar{E} = \lim_{T\to\infty} \frac{1}{T}\sum_{t = 1}^{T} \mathbbm{1}\left(E_{t}\neq 0\right) = \lim_{T\to\infty} \frac{1}{T}\sum_{t = 1}^{T} \mathbbm{1}\left(X_{t}\neq \hat{X}_{t}\right),
	\end{align}
 where $\mathbbm{1}(\cdot)$ is the indicator function. 
\item \textbf{Cost of Actuation Error:} this metric captures the significance of erroneous actions at the receiver side and how different errors may have diverse impact on the system performance. To study the cost of actuation error at time slot $t$, we denote $C_{i,j}$ the cost of error when the state of the source is $i$, i.e., $X_{t} = i$, and the state of the reconstructed source is $j\neq i$, i.e., $\hat{X}_{t}=j$. We assume that $C_{i,j}$ is fixed over time. Now, using $C_{i,j}$, we calculate the average cost of actuation error for an $N$-state DTMC model for the information source as
\begin{align}
	\label{Avg_Cost_ActuationError}
	\bar{C}_{A} = \sum_{i=0}^{N-1}\sum_{\substack{j=0 \\ j\neq i}}^{N-1} C_{i,j}\pi_{i,j},
\end{align}
where $\pi_{i,j}$ is the stationary distribution of the two-dimensional DTMC describing the joint status of the reconstructed source regarding the current state at the original source when the system is in an erroneous state, i.e., $\big(X_{t},\hat{X}_{t}\big)=\big(i,j\big), i\neq j$.

\item \textbf{Timing-Aware Error Metrics:} we introduce two new performance metrics of interest. There are several real-time and/or mission-critical applications where being consecutively in an erroneous state for some time may lead to safety issues or could even have catastrophic consequences for the system. For that, we propose the \emph{consecutive error} metric, which is defined as the number of consecutive time slots that the system is in an erroneous state. This metric captures the temporal sequence/evolution of errors, which can cumulatively have a serious impact on the actuation performance. We also propose a companion metric, coined \emph{cost of memory error}, which considers the memory of the actuation error when the system is not in a synced state over several consecutive time slots and it can be utilized as a penalty. We formally define these metrics in Section \ref{ConsecutiveTimeSlots}.
\end{enumerate}

\section{Real-time Reconstruction Error}
\label{RealTime}
In this section, we analyze the time-averaged reconstruction error under the randomized stationary policy and derive general expressions for the transition probabilities of $E_{t}$ for the two information source models (DTMC and BDMP). Since we lay here the theoretical foundation that we will also utilize in the other metrics, we devote a separate section.
We define the transition probabilities of $E_{t}$ as
\begin{align}
	\label{trans_prob}
	P_{i,j} = \mathrm{Pr}\big[E_{t+1} = j\big|E_{t} = i\big], \hspace{0.2 cm} \forall i,j\in\{0,1,\cdots, N-1\}.
\end{align}
 
where at time slot $t$ the synced state of the system is denoted by $E_{t} = 0$, while $E_{t}\neq 0$ denotes the system is in an erroneous state.

\subsection{Transition Probabilities using the DTMC information source model}
\label{NState1}
Using total probability theorem, one can write the transition probabilities $P_{i,j}$ given in \eqref{trans_prob} as

\begin{align}
\label{trans_prob2}
P_{i,j} \!&= \sum_{n=0}^{N-1} \mathrm{Pr}\big[E_{t+1} = j\big|E_{t} = i, X_{t} = n\big]\mathrm{Pr}\big[X_{t} = n\big|E_{t} = i\big]\notag\\
	\!&=\sum_{n=0}^{N-1} \frac{\mathrm{Pr}\big[E_{t+1} = j\big|E_{t} = i, X_{t} = n\big]\mathrm{Pr}\big[X_{t} = n, E_{t} = i\big]}{\mathrm{Pr}\big[E_{t}=i\big]}\notag\\
	\!&=\sum_{n=0}^{N-1}\!\Bigg(\!\frac{c_{1}\!\times\!\mathrm{Pr}\big[E_{t+1} = j\big|X_{t} = n, \hat{X}_{t}=n-i\big]\!\mathrm{Pr}\big[X_{t} = n, \hat{X}_{t}= n-i\big]}{\mathrm{Pr}\big[E_{t}=i\big]}\notag\\
	\!&+\! \frac{c_{2}\!\times\!\mathrm{Pr}\big[E_{t+1} = j\big|X_{t}= n,\hat{X}_{t}=n+i\big]\!\mathrm{Pr}\big[X_{t} = n, \hat{X}_{t}=n+i\big]}{\mathrm{Pr}\big[E_{t}=i\big]}\!\Bigg), \hspace{0.01 cm} \forall i,j\!\in\!\{0,1,\cdots, \!N\!-\!1\},
\end{align}
where $c_{1}$ and $c_{2}$ in \eqref{trans_prob2} are given by
\begin{align}
	\label{c1_c2}
	c_{1}=\mathbbm{1}\big(i=0\big)0.5+\mathbbm{1}\big(n\geqslant i, i\neq 0\big), \hspace{0.2cm}
c_{2}= \mathbbm{1}\big(i=0\big)0.5+\mathbbm{1}\big(n+i\leqslant N-1, i\neq 0\big).
\end{align}
Furthermore, the probability $\mathrm{Pr}\big[E_{t}=i\big]$ in \eqref{trans_prob2} can be written as
\begin{align}
\label{PEt_i}
\mathrm{Pr}\big[E_{t}=i\big] &= \sum_{m=0}^{N-1} \mathrm{Pr}\big[ X_{t} = m, E_{t}=i\big]\notag\\
	&=\sum_{m=0}^{N-1}\Bigg(f_{1}\times\mathrm{Pr}\big[X_{t} = m, \hat{X}_{t} =  m-i\big]+
	f_{2}\times\mathrm{Pr}\big[X_{t} = m, \hat{X}_{t} = m+i\big]\Bigg), 
\end{align}
where $f_{1}$ and $f_{2}$ in \eqref{PEt_i} are given by
\begin{align}
	\label{f1_f2}
f_{1}=\mathbbm{1}\big(i=0\big)0.5+\mathbbm{1}\big(m\geqslant i, i\neq 0\big), \hspace{0.2cm}
f_{2}= \mathbbm{1}\big(i=0\big)0.5+\mathbbm{1}\big(m+i\leqslant N-1, i\neq 0\big).
\end{align}
Note that the joint probability $\mathrm{Pr}\big[X_{t} =  m, \hat{X}_{t} = m\pm i\big]$ is the stationary distribution of the two-dimensional DTMC describing the joint status of the system regarding the current state at the original source. Now, using the natural evolution of the DTMC information source model,
the expression $\displaystyle\frac{\mathrm{Pr}\big[X_{t} = n, \hat{X}_{t} =n\pm i\big]}{\mathrm{Pr}\big[E_{t} = i\big]}$ given in \eqref{trans_prob2} can be simplified as

\begin{align}
	\label{Cond_Prob_X_E}
	\frac{\mathrm{Pr}\big[X_{t} = n, \hat{X}_{t} =n\pm i\big]}{\mathrm{Pr}\big[E_{t} = i\big]} = \mathbbm{1}\big(i=0\big)\frac{1}{N}+\mathbbm{1}\big(i\neq 0\big)\frac{1}{2(N-i)}.
\end{align}
To derive the conditional probability given in \eqref{trans_prob2}, we first define ${H}_{0}$ and ${H}_{1}$ as
\begin{align}
	\label{H0_H1}
{H}_{0} &= \mathrm{Pr}\Big[\alpha^{\text{s}}_{t+1}=1, \alpha^{\text{tx}}_{t+1}=1, h_{t+1} = 0\Big]=\mathrm{Pr}\Big[\alpha^{\text{s}}_{t+1}=1, \alpha^{\text{tx}}_{t+1}=1\Big]\mathrm{Pr}\Big[h_{t+1} = 0\Big]\notag\\
	{H}_{1} &= \mathrm{Pr}\Big[\alpha^{\text{s}}_{t+1}=1, \alpha^{\text{tx}}_{t+1}=1, h_{t+1} = 1\Big]=\mathrm{Pr}\Big[\alpha^{\text{s}}_{t+1}=1, \alpha^{\text{tx}}_{t+1}=1\Big]\mathrm{Pr}\Big[h_{t+1} = 1\Big],
\end{align}
where ${H}_{0}$ and ${H}_{1}$ are the joint probability density function (PDF) of sampling and transmissions actions at time slot $t+1$ when we have failed and successful transmission, respectively. We define $\mathrm{Pr}\{\alpha^{\text{s}}_{t+1} = 0\}=1-\mathrm{Pr}\{\alpha^{\text{s}}_{t+1} = 1\}$ as the probability that the source is not sampled at time slot $t+1$. In this work, we assume that the transmitter can send samples immediately after the sampler performs sampling. This means that the joint probability of sampling and transmissions actions at time slot $t+1$ when the sampler performs sampling is equal to $p_{\alpha^{\text{s}}} = \mathrm{Pr}\big[\alpha^{\text{s}}_{t+1}=1\big]$. Therefore, \eqref{H0_H1} can be simplified as
\begin{equation}
	\label{H0_H1_2}
	{H}_{0} = p_{\alpha^{\text{s}}}\big(1-p_{s}\big),
	{H}_{1} = p_{\alpha^{\text{s}}}p_{s},
\end{equation}
where $p_{s}$ is the success probability given by \eqref{success_Prob}. Now, we obtain the general expressions for the transition probabilities.
\begin{lemma}
	\label{lemma_P00}
The general expression for $P_{0,0}$ can be expressed as
\begin{align}
	\label{P00_1}
	P_{0,0} =  q+(N-1)p{H}_{1}.
\end{align}
	Proof. See Appendix \ref{Appendixlemma1}.
\end{lemma}
Similar to Lemma \ref{lemma_P00}, we can obtain the transition probabilities $P_{i,j}$ for different values of $i$ and $j$ as follows
\begin{align}
	\label{pij_M1_1}
	P_{i,0}&=\mathbbm{1}\big(1\leqslant i\leqslant N-1\big)\Big(p+q{H}_{1}+(N-2)p{H}_{1}\Big).\notag\\
	P_{0,j} &=\mathbbm{1}\big(1\leqslant j\leqslant N-1\big)\bigg[2\Big(1-\frac{j}{N}\Big)\Big(p{H}_{0}+p\big(1-p_{\alpha^{\text{s}}}\big)\Big)\bigg].\notag\\
	P_{i,i} &\!=
	\begin{cases}
	    \frac{N-2i}{N-i}\Big(p{H}_{0}+p\big(1-p_{\alpha^{\text{s}}}\big)\Big)+q{H}_{0}+q\big(1-p_{\alpha^{\text{s}}}\big), \hspace{0.2cm} 1\leqslant i \leqslant \frac{N-1}{2}\\
		q{H}_{0}+q\big(1-p_{\alpha^{\text{s}}}\big), \hspace{0.1cm} \frac{N}{2}\leqslant i\leqslant N-1\\
		0, \hspace{0.1cm} i\geqslant N.
	\end{cases}\notag\\
	P_{1,j} &=\mathbbm{1}\big(2\leqslant j\leqslant N-1\big)\bigg[\frac{2N-2j-1}{N-1}\Big(p{H}_{0}+p\big(1-p_{\alpha^{\text{s}}}\big)\Big)\bigg].\notag\\
	P_{i,1} & = \mathbbm{1}\big(2\leqslant i\leqslant N-1\big)\bigg[\frac{2N-2i-1}{N-i}\Big(p{H}_{0}+p\big(1-p_{\alpha^{\text{s}}}\big)\Big)\bigg].
\end{align}
For $i\geqslant 2$, and $j>i$,  $P_{i,j}$ is given by
\begin{align}
	\label{pij_M1_3}
	P_{i,j} \!\!&=\!\!
	\begin{cases}
		\!\frac{N-j}{N-i}\Big(\!p{H}_{0}\!+\! p\big(1-p_{\alpha^{\text{s}}}\big)\!\Big)
		, \hspace{0.05cm}j\!+\!1\!\leqslant\! N\! \leqslant\! i\!+\!j\!-\!1\\
		\!\frac{2N-i-2j}{N-i}\Big(p{H}_{0}+p\big(1-p_{\alpha^{\text{s}}}\big)\Big),\hspace{0.05cm}N\!\geqslant\! i+j\\
		0,\hspace{0.05cm} N\leqslant j.
	\end{cases}
\end{align}
For $j\geqslant 2$, and $i>j$, $P_{i,j}$ is given by
\begin{align}
	\label{pij_M1_4}
	P_{i,j} &=
	\begin{cases}
		p{H}_{0}+p\big(1-p_{\alpha^{\text{s}}}\big),\hspace{0.1cm} i+1\leqslant N \leqslant i+j-1\\
		\frac{2N-j-2i}{N-i}\Big(p{H}_{0}+p\big(1-p_{\alpha^{\text{s}}}\big)\Big),\hspace{0.1cm} N \geqslant i+j\\
		0,\hspace{0.05cm} N\leqslant i.
	\end{cases}
\end{align}
\subsection{Transition Probabilities using the BDMP information source model}
\label{NState2}
Using the expression given in \eqref{trans_prob2}, we can write the transition probabilities $P_{i,j}$ as
\begin{align}
	\label{trans_prob3}
	P_{i,j}
	\!&=\sum_{n=0}^{N-1} \Bigg(\frac{c_{1}\times\mathrm{Pr}\big[E_{t+1} = j\big|X_{t} = n, \hat{X}_{t}=n-i\big]\mathrm{Pr}\big[X_{t} = n, \hat{X}_{t}= n-i\big]}{\mathrm{Pr}\big[E_{t}=i\big]}\notag\\
	\!&+ \frac{c_{2}\!\times\!\mathrm{Pr}\big[E_{t+1} = j\big|X_{t}= n,\hat{X}_{t}=n+i\big]\!\mathrm{Pr}\big[X_{t} = n, \hat{X}_{t}=n+i\big]}{\mathrm{Pr}\big[E_{t}=i\big]}\!\Bigg), \hspace{0.01 cm} \forall i,j\!\in\!\{0,1,\cdots, \!N\!-\!1\},
\end{align}
where $c_{1}$ and $c_{2}$ are given by \eqref{c1_c2}. Furthermore, the probability $\mathrm{Pr}\big[E_{t}=i\big]$ is given by \eqref{PEt_i}.

In what follows, we assume that $\pi_{n,n\pm i} = \mathrm{Pr}\big[X_{t} = n, \hat{X}_{t} = n\pm i\big]$, where $\pi_{n,n\pm i}$ is the stationary distribution of the two-dimensional DTMC describing the joint status of the system regarding the current state at the original source. In Appendix \ref{Appendixlemma3}, we consider a two- and a three-state BDMP information source and obtain $\pi_{n,n\pm i}$. Then, one can utilize the same procedure to obtain $\pi_{n,n\pm i}$ for the $N$-state BDMP source for different values of $n$ and $i$. Now, using \eqref{trans_prob3}, and \eqref{c1_c2}, we can calculate $P_{0,0}$ as

\begin{align}
\label{P00_M2}
P_{0,0} &=\frac{1}{	\mathrm{Pr}\big[E_{t}=0\big]}\Bigg(\mathrm{Pr}\big[E_{t+1} = 0\big|X_{t} = 0, \hat{X}_{t} = 0\big]\mathrm{Pr}\big[X_{t} = 0, \hat{X}_{t} = 0\big]
\notag\\&+\sum_{n=1}^{N-2}\mathrm{Pr}\big[E_{t+1} = 0\big|X_{t} = n, \hat{X}_{t} = n\big]\mathrm{Pr}\big[X_{t} = n, \hat{X}_{t} = n\big]\notag\\&+\mathrm{Pr}\big[E_{t+1} = 0\big|X_{t} = N-1, \hat{X}_{t} = N-1\big]\mathrm{Pr}\big[X_{t} = N-1, \hat{X}_{t} = N-1\big]\Bigg).
\end{align}
 Using a similar procedure as in Lemma \ref{lemma_P00}, $P_{0,0}$ can be written as
 \begin{align}
 	\label{P00_M22}
 	P_{0,0} \!=\! \frac{1}{\mathrm{Pr}\big[E_{t}=0\big]}\!\Bigg[\!\Big(\!1\!-\!p\!+\!p{H}_{1}\!\Big)\pi_{0,0}\!\!+\!\!\sum_{n=1}^{N-2}\!\Big(\!1\!-\!p\!-\!q\!+\!(p+q){H}_{1}\!\Big)\pi_{n,n}\!\!+\!\!\Big(\!1\!-\!q\!+\!q{H}_{1}\!\Big)\pi_{N\!-\!1,N\!-\!1}\!\Bigg].
 \end{align}
We can derive the transition probabilities $P_{i,j}$ for different values of $i$ and $j$ for the BDMP describing the information source as follows
\begin{align}
P_{0,i} &=\frac{\mathbbm{1}\big(i=1\big)}{	\mathrm{Pr}\big[E_{t}=0\big]}\Bigg[\Big(p{H}_{0}+p\big(1-p_{\alpha^{\text{s}}}\big)\Big)\pi_{0,0}+\displaystyle\sum_{n=1}^{N-2}\Big((p+q){H}_{0}+(p+q)\big(1-p_{\alpha^{\text{s}}}\big)\Big)\pi_{n,n}\notag\\
&\hspace{5cm}+\Big(q{H}_{0}+q\big(1-p_{\alpha^{\text{s}}}\big)\Big)\pi_{N-1,N-1}\Bigg],\notag\\
P_{1,0}&=\frac{1}{\mathrm{Pr}\big[E_{t}=1\big]}\!\Bigg[\!\sum_{n=1}^{N-2}\!\bigg(\!\Big(q\!+\!(1-p-q){H}_{1}\!+\!\!p{H}_{1}\Big)\pi_{n,n-1}\!+\!\!\Big(p\!+\!(1-p-q){H}_{1}\!+\!q{H}_{1}\!\Big)\pi_{n,n+1}\!\bigg)\notag\\
&\hspace{5cm}+\Big(p+(1-p){H}_{1}\Big)\pi_{0,1}+\Big(q+(1-q){H}_{1}\Big)\pi_{N-1,N-2}\Bigg]\notag
\end{align}
\begin{align}
\label{pij_M2_1}
P_{i,0} & = 0, \hspace{0.1cm} \forall i\geqslant N.\\ 
P_{i,0} &=\frac{\mathbbm{1}\big(2\leqslant i\leqslant N-1\big)}{\mathrm{Pr}\big[E_{t}=i\big]}\Bigg[\!\Big((1-p){H}_{1}\!+\!p{H}_{1}\Big)\pi_{0,i}+\Big((1-q){H}_{1}+q{H}_{1}\Big)\pi_{N-1,N-1-i}+{F}_{1}\!\Bigg],
\end{align}
where $F_{1}$ can be obtained as
\begin{align}
\label{F1}
    {F}_{1} = \mathbbm{1}\big(N\geqslant i+2\big)\sum_{n=1}^{N-2}\Big((p+q){H}_{1}+(1-p-q){H}_{1}\Big)\Big(\alpha_{1}\pi_{n,n-i}+\alpha_{2}\pi_{n,n+i}\Big),
\end{align}
where $\alpha_{1}$ and $\alpha_{2}$ in \eqref{F1} are given by
\begin{align}
    \alpha_{1} = \mathbbm{1}\big(n-i\geqslant 0\big), \hspace{0.2cm} \alpha_{2} = \mathbbm{1}\big(n+i\leqslant N-1\big).
    \label{ab}
\end{align}
In what follows, we obtain $P_{i,j}$ when $i$ and $j$ are not equal to $0$.
\begin{align}
		P_{i,j} &= \mathbbm{1}\big(j=i-1\big)P_{i,i-1}+\mathbbm{1}\big(j=i+1\big)P_{i,i+1}+\mathbbm{1}\big(j=i\big)P_{i,i},
\end{align}
where $P_{i,i-1}$, $P_{i,i+1}$ and $P_{i,i}$ are given by
\begin{align}
\label{Pijmp}
P_{i,i-1}&=\frac{\mathbbm{1}\big(2\leqslant i\leqslant N\!-\!1\big)}{\mathrm{Pr}\big[E_{t}=i\big]}\!\!\sum_{n=0}^{N-1}\!\Big[\alpha_{1} \Big(q{H}_{0}\!+\!q\big(1-p_{\alpha^{\text{s}}}\big)\Big)\pi_{n,n-i}\!\!+\!\!\alpha_{2} \Big(p{H}_{0}\!+\!p\big(1-p_{\alpha^{\text{s}}}\big)\Big)\pi_{n,n+i}\Big],\notag\\
P_{i,i+1}&= \frac{\mathbbm{1}\big(1\leqslant i \leqslant N\!-\!2\big)}{\mathrm{Pr}\big[E_{t}=i\big]}\!\!\sum_{n=1}^{N-2}\!\Big[\alpha_{1} \Big(p{H}_{0}\!+\!p\big(1-p_{\alpha^{\text{s}}}\big)\Big)\pi_{n,n-i}\!\!+\!\!\alpha_{2} \Big(q{H}_{0}\!+\!q\big(1-p_{\alpha^{\text{s}}}\big)\Big)\pi_{n,n+i}\Big],\notag\\
P_{i,i} &=\frac{\mathbbm{1}\big(1\leqslant i \leqslant N-1\big)}{\mathrm{Pr}\big[E_{t}=i\big]}\Bigg[\Big((1-p){H}_{0}+(1-p)\big(1-p_{\alpha^{\text{s}}}\big)\Big)\pi_{0,i}\notag\\&\hspace{2.9cm}+\Big((1-q){H}_{0}+(1-q)\big(1-p_{\alpha^{\text{s}}}\big)\Big)\pi_{N-1,N-1-i}+{F}_{2}\Bigg],    
\end{align}
where ${F}_{2}$ in \eqref{Pijmp} is given by
\begin{align}
	\label{pij_M2_1}
{F}_{2} =\mathbbm{1}\big(N\geqslant i+2\big)\sum_{n=1}^{N-2}\Big((1-p-q){H}_{0}+(1-p-q)\big(1-p_{\alpha^{\text{s}}}\big)\Big)\Big(\alpha_{1}\pi_{n,n-i}+\alpha_{2}\pi_{n,n+i}\Big),
\end{align}
where $\alpha_{1}$ and $\alpha_{2}$ in \eqref{Pijmp} and \eqref{pij_M2_1} are given by \eqref{ab}.
Using the transition probabilities given in subsection \ref{NState1} and \ref{NState2}, the probability that the system is in an erroneous state, $P_{E}$, or the time-averaged reconstruction error can be derived by obtaining the state stationary distributions of the transition matrix. As an example, we calculate $P_{E}$ for $N=3$
\begin{equation}
	\label{PE}
	P_{E} = \frac{{\Phi}}{\Phi+P_{2,0}-P_{2,0}P_{1,1}+P_{1,0}P_{2,1}},
\end{equation}
where $\Phi =1+P_{2,1}-P_{1,1}-P_{0,0}-P_{0,0}P_{2,1}+P_{0,0}P_{1,1}+P_{0,1}P_{2,0}-P_{0,1}P_{1,0}$.

Using \eqref{timeAvg_Error}, we define the variance of real-time reconstruction error for an observation interval $[1, T]$ as
\begin{align}
	\label{var_Error}
		\mathrm{Var}(E_{t}) &\!=\! \lim_{T\to\infty} \!\frac{1}{T}\sum_{t = 1}^{T} \!\mathbbm{1} (E_{t}\neq 0)^{2}\!\!-\!\!\left(\!\lim_{T\to\infty} \frac{1}{T}\sum_{t = 1}^{T} \mathbbm{1}(E_{t}\neq 0)\!\right)^{2}\notag\\& = \lim_{T\to\infty} \frac{1}{T}\sum_{t = 1}^{T} \mathbbm{1}\left(X_{t}\neq \hat{X}_{t}\right)^{2}-\left(\lim_{T\to\infty} \frac{1}{T}\sum_{t = 1}^{T}\mathbbm{1} \left(X_{t}\neq \hat{X}_{t}\right)\right)^{2}\!=P_{E}-P_{E}^{2}.
\end{align}
\begin{remark}
We can analytically prove that for a three-state DTMC, the randomized stationary policy  has higher time-averaged reconstruction error for $p_{\alpha^{\text{s}}} < 1$ compared to the semantics-aware policy, while it has lower time-averaged reconstruction error in comparison with the change-aware policy only if $p_{\alpha^{\text{s}}}\geqslant \frac{2p}{1-p_{\text{s}}(1-2p)}$. Furthermore, the randomized stationary policy has lower reconstruction error variance for $0<p_{\alpha^{\text{s}}}\leqslant\frac{pp_{\text{s}}-3p^{2}(1-p_{\text{s}})}{(2p+3p^2-1)p^{2}_{\text{s}}-(p+3p^{2})p_{\text{s}}}$ compared to semantics-aware policy, and for $\frac{p(3+p_{\text{s}})}{pp_{\text{s}}(3+p_{\text{s}})-2p_{\text{s}}}\leqslant p_{\alpha^{\text{s}}} <1$ in comparison with the change-aware policy.
\end{remark}
\section{Analysis for Cost of Actuation Error and Timing-aware Error Metrics}
\label{Analyze_Cost_and_ConsecutiveTimeSlots}
\par In this section, we calculate the average cost of actuation error under the randomized stationary policy. Then, we analyze the timing-aware error metrics, including the average consecutive error and the cost of memory error.
\subsection{Cost of Actuation Error}
\label{Cost_Actutation_Error}
\par As we discussed in Section \ref{metrics}, to obtain the average cost of actuation error we need to calculate $\pi_{i,j}$. In the following, we derive $\pi_{i,j}$ using two-state DTMC and BDMP models for describing the source evolution. Then, one can utilize the same procedure to obtain $\pi_{i,j}$ for $N$-state DTMC and BDMP information sources.
\begin{lemma}
	\label{lemma_PI_ij_DTMCmodel}
	For a two-state DTMC information source model, the stationary distribution $\pi_{i,j}$ for the randomized stationary policy is given by
	\begin{align}
		\label{pij_M1_N2}
		\pi_{0,0}&=\frac{p+(1-p)p_{\alpha^{\text{s}}}p_{\text{s}}}{4p+2p_{\alpha^{\text{s}}}p_{\text{s}}-4pp_{\alpha^{\text{s}}}p_{\text{s}}},\hspace{0.2cm}
		\pi_{0,1}=\frac{p(1-p_{\alpha^{\text{s}}}p_{\text{s}})}{4p+2p_{\alpha^{\text{s}}}p_{\text{s}}-4pp_{\alpha^{\text{s}}}p_{\text{s}}}\notag\\
		\pi_{1,0}&=\frac{p(1-p_{\alpha^{\text{s}}}p_{\text{s}})}{4p+2p_{\alpha^{\text{s}}}p_{\text{s}}-4pp_{\alpha^{\text{s}}}p_{\text{s}}},\hspace{0.2cm}
		\pi_{1,1}=\frac{p+(1-p)p_{\alpha^{\text{s}}}p_{\text{s}}}{4p+2p_{\alpha^{\text{s}}}p_{\text{s}}-4pp_{\alpha^{\text{s}}}p_{\text{s}}}.
	\end{align}
	Proof. See Appendix \ref{Appendixlemma2}.
\end{lemma}
\begin{lemma}
	\label{lemma_PI_ij_BirthDeathModel}
	For a two-state the BDMP information source model, the stationary distribution $\pi_{i,j}$ for the randomized stationary policy is given by
\begin{align}
	\label{pij_M2_N2}
	\pi_{0,0} &=\frac{q\big[q+(1-q)p_{\alpha^{\text{s}}}p_{\text{s}}\big]}{(p+q)\big[p(1-p_{\alpha^{\text{s}}}p_{\text{s}})\!+\!q\!+\!(1-q)p_{\alpha^{\text{s}}}p_{\text{s}}\big]},\hspace{0.1cm}
	\pi_{0,1} =\frac{pq(1-p_{\alpha^{\text{s}}}p_{\text{s}})}{(p+q)\big[p(1-p_{\alpha^{\text{s}}}p_{\text{s}})\!+\!q\!+\!(1-q)p_{\alpha^{\text{s}}}p_{\text{s}}\big]}\notag\\
	\pi_{1,0} &=\frac{pq(1-p_{\alpha^{\text{s}}}p_{\text{s}})}{(p+q)\big[p(1-p_{\alpha^{\text{s}}}p_{\text{s}})\!+\!q\!+\!(1-q)p_{\alpha^{\text{s}}}p_{\text{s}}\big]},\hspace{0.1cm}
	\pi_{1,1} =\frac{p\big[p+(1-p)p_{\alpha^{\text{s}}}p_{\text{s}}\big]}{(p+q)\big[p(1-p_{\alpha^{\text{s}}}p_{\text{s}})\!+\!q\!+\!(1-q)p_{\alpha^{\text{s}}}p_{\text{s}}\big]}.
\end{align}
	Proof. See Appendix \ref{Appendixlemma3}.
\end{lemma}
\subsection{Timing-Aware Error Metrics}
\label{ConsecutiveTimeSlots}
\par Our objective is to obtain the \textit{average consecutive error} and the \textit{cost of memory error} as a means to evaluate the performance. To this end, we describe the evolution of the state of consecutive error by a Markov Chain as illustrated in Fig. \ref{Consequtive Error}. In this DTMC model, the synced state is denoted by $0$, whereas $i\geqslant 1$ denotes time slot during which the system is in an erroneous state. The state stationary distribution of this DTMC model can be obtained as
\begin{align}
	\label{Pn_ConsecutiveError}
	\pi_{0} = 1-P_{E}, 
	\pi_{n}  = P^{n}_{E} \pi_{0} = (1-P_{E})P^{n}_{E},
\end{align}
where $\pi_{0}$ is the probability the system is in synced state, and $\pi_{n}$ is the probability the system is in an erroneous state for $n$ consecutive time slots. $P_{E}$ is the time-averaged reconstruction error, which is given by \eqref{PE} for $N=3$. Now, using \eqref{Pn_ConsecutiveError}, the \textit{average consecutive error} can be written as
\begin{align}
    \label{Avg_Cons_error}
    \bar{C}_{E} = \sum_{x = 1}^{\infty} x\pi_{x} = (1-P_{E})\sum_{x = 1}^{\infty} x P^{x}_{E} = \frac{P_{E}}{1-P_{E}}.
\end{align}
We define the \textit{memory of actuation error} as
\begin{align}
	\label{memoryerror}
	C_{\text{M}}(x) = 
	\begin{cases}
		0, &x = 0,\\
		\kappa^{x}, &x = 1, 2 ,\cdots, n,
	\end{cases}
\end{align}
where $n$ is finite. Using $C_{\text{M}}(x)$, we can define the cost of memory error over $n$ consecutive time slots as
\begin{align}
	\label{Avg_ConsecutiveError}
	\bar{C}^{\text{M}}_{E} = \sum_{x = 1}^{n} C_{\text{M}}(x)\pi_{x} = \sum_{x=1}^{n}\kappa^{x}P_{E}^{x}(1-P_E) = (1-P_E)\kappa P_E\bigg[\frac{1-(\kappa P_{E})^{n}}{1-\kappa P_{E}}\bigg].
\end{align}
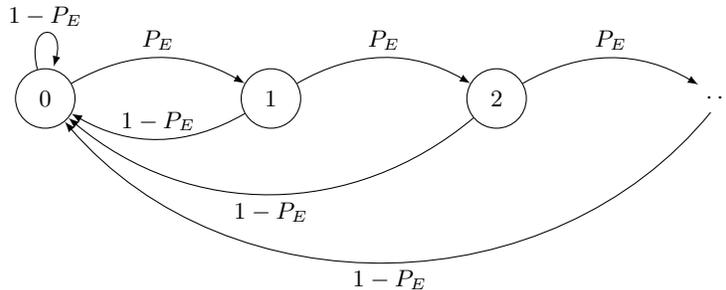
\begin{figure}[!t]
	\centering
		\begin{tikzpicture}[start chain=going left,->,>=latex,node distance=3cm,on grid,auto]
\footnotesize
\node[on chain]                        (g) {$\cdots$};
		\node[state, on chain]                 (3) {$2$};
		\node[state, on chain]                 (2) {$1$};
		\node[state, on chain]                 (1) {$0$};
		\draw[>=latex]
		(1)   edge[loop above] node {$1-P_{E}$}   (1)
		(1) edge  [bend left=30] node {$P_{E}$} (2)
		(2) edge  [bend left=30] node {$P_{E}$} (3)
		(2) edge  [bend left=30] node[above] {$1-P_{E}$} (1)
		(3) edge  [bend left=40] node {$1-P_{E}$} (1)
		(3) edge  [bend left=30] node {$P_{E}$} (g)
		(g) edge  [bend left=50] node {$1-P_{E}$} (1)
		;
	\end{tikzpicture}
	\vspace*{1ex}
	\caption{DTMC describing the state of consecutive error.}
	\label{Consequtive Error}
\end{figure}
\section{Sampling Policy Optimization}
\label{Optimization_Problem}
\par In this section, we formulate and solve two optimization problems. In the first one, presented in Section \ref{OptimizationProb1}, we find an optimal randomized stationary policy to minimize the time-averaged reconstruction error, while keeping the time-averaged sampling cost under a given threshold. In the second one, presented in Section \ref{OptimizationProblem2}, the objective is to find an optimal sampling policy that minimizes the average consecutive error with a constraint on the time-averaged sampling cost being less than a certain threshold. 

\subsection{Minimizing the time-averaged reconstruction error}
\label{OptimizationProb1}
\par The objective of this optimization problem is to find the optimal value of the joint probability of sampling and transmission, $p_{\alpha^{\text{s}}}=\mathrm{Pr}\{\alpha^{\text{s}}_{t+1}=1\}$, so as to minimize the time-averaged reconstruction error. Let $\delta$ and $\delta_{\text{max}}$ are strictly positive real values, representing the cost of sampling at each attempted transmission and the total average sampling cost, respectively. We formulate the optimization problem as
\begin{subequations}
\label{Optimization_problem1}
\begin{align}
&\underset{p_{\alpha^{\text{s}}}}{\text{minimize}}\hspace{0.3cm}\hspace{0.3cm} P_{E}\\
&\text{subject to}\hspace{0.3cm} \lim_{T \to \infty}\frac{1}{T}\sum_{t=1}^{T}\delta \mathbbm{1}\{\alpha^{\text{s}}_{t}=1\} \leqslant\delta_{\text{max}},\label{Optimization_prob1_constraint}
\end{align}
\end{subequations}
and we define $\eta = \frac{\delta_{\text{max}}}{\delta}\leqslant 1$. The constraint in \eqref{Optimization_prob1_constraint} is the time-averaged sampling cost which  can be simplified as
\begin{align}
\label{Const_1}
    p_{\alpha^{\text{s}}}\leqslant \eta.
\end{align}
To solve this optimization problem we first consider a two-state DTMC information source; this can be easily extended to an $N$-state DTMC with $N>2$. 
Using \eqref{pij_M1_N2}, the time-averaged reconstruction error can be calculated as
\begin{align}
	P_{E} = \pi_{0,1}+\pi_{1,0} = \frac{2\big(p-p p_{\alpha^{\text{s}}} p_{\text{s}}\big)}{4p+2p_{\alpha^{\text{s}}}p_{\text{s}}-4pp_{\alpha^{\text{s}}}p_{\text{s}}}\label{PE_M1_N2}.
\end{align}
Now, using \eqref{Const_1} and \eqref{PE_M1_N2}, the optimization problem can be formulated as follows
\begin{subequations}
\label{Optimization_problem1_1}
\begin{align}
&\underset{p_{\alpha^{\text{s}}}}{\text{minimize}}\hspace{0.3cm} \frac{2\big(p-p p_{\alpha^{\text{s}}} p_{\text{s}}\big)}{4p+2p_{\alpha^{\text{s}}}p_{\text{s}}-4pp_{\alpha^{\text{s}}}p_{\text{s}}}\label{PE_M1_N2_2}
\\
&\text{subject to}\hspace{0.3cm} p_{\alpha^{\text{s}}}\leqslant \eta.\label{Optimization_problem1_constraint}
\end{align}
\end{subequations}
\begin{lemma}
	\label{Optimization1_lemma1}
The minimum value of the optimization problem in \eqref{Optimization_problem1_1}, $P^{*}_{E}$, is obtained as
\begin{align}
	\label{PE_min}
	P^{*}_{E} =  \frac{2\big(p-p  p_{\text{s}}\eta\big)}{4p+2p_{\text{s}}\eta-4pp_{\text{s}}\eta}.
\end{align}
Proof. See Appendix \ref{Appendix_Optimization1_Lemma1}.
\end{lemma}

We now consider the BDMP model to derive the objective function of the optimization problem. Using \eqref{pij_M2_N2}, we can obtain the time-averaged reconstruction error for a two-state BDMP information source when the sampler performs sampling with the probability $p_{\alpha^{\text{s}}}$ and when the sampler never performs sampling as
\begin{align}
	P_{E} &=\pi_{0,1}+\pi_{1,0}=\displaystyle\frac{2pq\big(1-p_{\alpha^{\text{s}}}p_{\text{s}}\big)}{(p+q)\Big[p\big(1-p_{\alpha^{\text{s}}}p_{\text{s}}\big)+(1-q)p_{\alpha^{\text{s}}}p_{\text{s}}+q\Big]},\hspace{0.2cm} \mathrm{Pr}\{\alpha^{\text{s}}_{t+1} = 1\}\neq 0\label{PEM2}\\
	P^{\text{NS}}_{E}&=\mathbbm{1}\left(\mathrm{Pr}\{\alpha^{\text{s}}_{t+1} = 1\}=0, \hat{X}_{0} = 0\right)\frac{p}{p+q}+\mathbbm{1}\left(\mathrm{Pr}\{\alpha^{\text{s}}_{t+1} = 1\}=0, \hat{X}_{0} = 1\right)\frac{q}{p+q},
 \label{PEM2_NS}
\end{align}
where $\hat{X}_{0}=0$ $(\text{or}\hspace{0.1cm} \hat{X}_{0}=1)$ is the condition that the receiver at $t=0$ is at the state 0 $(\text{or}\hspace{0.1cm}1)$.
According to \eqref{PEM2} and \eqref{PEM2_NS}, $P_{E}$ is smaller than $P^{\text{NS}}_{E}$ when
\begin{align}
    p_{\alpha^{\text{s}}}>\frac{q-p}{p_{\text{s}}(1+q-p)}, \hspace{0.2cm} \hat{X}_{0}=0\label{Const_M2_1}\\
    p_{\alpha^{\text{s}}}>\frac{p-q}{p_{\text{s}}(1+p-q)}, \hspace{0.2cm} \hat{X}_{0}=1\label{Const_M2_2}.
\end{align}
To solve this optimization problem we first consider that the receiver at $t = 0$ is at the state $0$, that is, $\hat{X}_{0} = 0$. According to this expression, when $q\leqslant p$, we can always find a probability $p_{\alpha^{\text{s}}}\in [0, 1]$ that satisfies the condition given in eq. \eqref{Const_M2_1}. Since  $P_{E}$ given in \eqref{PEM2} is decreasing in $p_{\alpha^{\text{s}}}$, the optimal value of $p_{\alpha^{\text{s}}}$ that minimizes $P_{E}$ is given by $p^{*}_{\alpha^{\text{s}}} = \eta$. Now, we consider the case when $q>p$. When $p_{\text{s}} < \displaystyle \frac{q-p}{1+q-p}$, we cannot find a probability $p_{\alpha^{\text{s}}}\in [0, 1]$ that satisfies the condition given in \eqref{Const_M2_1}. Therefore, in that case, it is optimal to not perform sampling. Now, we assume that $\displaystyle p_{\text{s}}> \frac{q-p}{1+q-p}$. In this case, the optimal sampling policy is to perform sampling with probability $\eta$ if and only if $\displaystyle\eta>\frac{q-p}{ p_{\text{s}}(1+q-p)}$. Otherwise, the optimal policy is to not perform sampling. When $\displaystyle\eta>\frac{q-p}{p_{\text{s}}(1+q-p)}$, the optimal value of $p_{\alpha^{\text{s}}}$ and the minimum value of time-averaged reconstruction error are given by
\begin{align}
    \label{Optimal_M2}
    p^{*}_{\alpha^{\text{s}}} = \eta,\hspace{0.3cm}
    P^{*}_{E} =\displaystyle\frac{2pq\big(1-p_{\text{s}}\eta\big)}{(p+q)\Big[p\big(1-p_{\text{s}}\eta\big)+(1-q)p_{\text{s}}\eta+q\Big]}.
\end{align}
Now, we consider that the receiver at $t=0$ is at the state 1. In that case, we can similarly prove that in the following two cases, the optimal sampling policy is to perform sampling with probability $p^{*}_{\alpha^{\text{s}}}=\eta$. First, when $p\leqslant q$, and second, when $p>q$ and we have $\displaystyle p_{\text{s}}> \frac{p-q}{1+p-q}$ and $\displaystyle\eta>\frac{p-q}{p_{\text{s}}(1+p-q)}$. Otherwise, it is optimal to refrain from performing sampling.

\subsection{Minimizing the average consecutive error}
\label{OptimizationProblem2}
\par
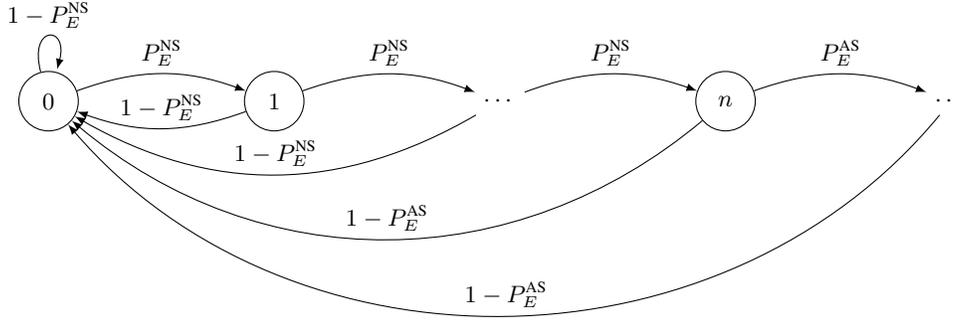
\begin{figure}[!t]
	\centering
	\begin{tikzpicture}[start chain=going left,->,>=latex,node distance=3cm,on grid,auto]
 \footnotesize
		\node[on chain]                        (h2) {$\cdots$};
	   \node[state, on chain]              (h1) {$n$};
		\node[on chain]               (g) {$\cdots$};
		\node[state, on chain]                 (2) {$1$};
		\node[state, on chain]                 (1) {$0$};
		\draw[>=latex]
		(1)   edge[loop above] node {$1-P^{\text{NS}}_{E}$}   (1)
		(1) edge  [bend left=20] node {$P^{\text{NS}}_{E}$} (2)
		(2) edge  [bend left=20] node[above] {$1-P^{\text{NS}}_{E}$} (1)
        (2) edge  [bend left=20] node[above] {$P^{\text{NS}}_{E}$} (g)
		(g) edge  [bend left=30] node[above] {$1-P^{\text{NS}}_{E}$} (1)
		(g) edge  [bend left=20] node[above] {$P^{\text{NS}}_{E}$} (h1)
		(h1) edge  [bend left=40] node[above] {$1-P^{\text{AS}}_{E}$} (1)
		(h1) edge  [bend left=20] node[above] {$P^{\text{AS}}_{E}$} (h2)
		(h2) edge  [bend left=50] node[above] {$1-P^{\text{AS}}_{E}$} (1)
		;
	\end{tikzpicture}
	\vspace*{-5ex}
	\caption{DTMC model for the evolution of the state of consecutive error.}
	\label{Fig_ConsecutiveError2}
\end{figure}

 The objective of the second optimization problem is to find an optimal sampling policy that minimizes the average consecutive error under a constraint on the time-averaged sampling cost. Here, it is assumed that each attempted sampling occurs a sampling cost $\delta$, and we require that the time-averaged sampling cost does not exceed a given threshold $\delta_{\text{max}}$. Evidently, in order to satisfy this constraint, the sampler cannot perform sampling at each time slot. Therefore, we propose a \emph{wait-then-generate} sampling policy, under which it is important that the sampler chooses when to perform sampling so as to minimize the average consecutive error while satisfying the sampling cost constraint. To this end, we formulate the optimization problem as follows
\begin{subequations}
	\label{OptimizationProb2}
\begin{align}
&\underset{n}{\text{minimize}}\hspace{0.3cm}\bar{C}_{E}(n)\label{OptimizationProb2_Objfunc}\\
&\text{subject to}\hspace{0.3cm} \frac{1}{T}\sum_{t=1}^{T}\delta \mathbbm{1}\{\alpha^{\text{s}}_{t}=1\} \leqslant\delta_{\text{max}},\label{Optimization_problem2_constraint}
\end{align}
\end{subequations}
where $n\in \mathbb{N}$, and it is assumed that for any time slot $t<n$ the sampler remain idle, otherwise it performs sampling. Moreover, $\bar{C}_{E}(n)$ is the infinite horizon average of consecutive error for that $n$. Now, we can model the evolution of the consecutive error as a DTMC, as shown in Fig. \ref{Fig_ConsecutiveError2}.
Therein, $P^{\text{NS}}_{E}$ is the time-averaged reconstruction error when no sampling is performed, which is equal to $1/2$ for the DTMC model and is given by \eqref{PEM2_NS} for the BDMP information source model. Moreover, $P^{\text{AS}}_{E}$ is the time-averaged reconstruction error when the sampler always performs sampling, which can be obtained when $p_{\alpha^{\text{s}}} = 1$. 
\begin{lemma}
	\label{Optimization2_lemma1}
	The optimal value of $n$ that minimizes $\bar{C}_{E}(n)$ in \eqref{OptimizationProb2} is given by
\begin{align}
	\label{n_optimal}
	n^{*} = \left\lceil{\log_{P^{\text{NS}}_{E}}\left(\frac{\eta\big(1-P^{\text{AS}}_{E}\big)}{1-(1-\eta)P^{\text{NS}}_{E}-\eta P^{\text{AS}}_{E}}\right)}\right\rceil.
\end{align}
	Proof. See Appendix \ref{Appendix_Optimization2_Lemma1}.
\end{lemma}
\begin{remark}
In what follows, RS policy is the abbreviation for randomized stationary policy, while RSC policy refers to the randomized stationary policy in the constrained optimization problem.
\end{remark}
\section{Simulation Results}
\label{Simulation_Results}
\par In this section, we validate our analysis and assess the performance of the sampling policies in terms of time-averaged reconstruction error and cost of memory error. Simulation results are obtained averaging over $10^{7}$ time slots and the parameters are set to $r = 30 \hspace{.05cm}\text{m}$, $\sigma^{2} = -100\hspace{.05cm}\text{dBm}$, $\alpha = 4$, $P_{\text{tx}} = 1 \hspace{.05cm}\text{mW}$. We also consider two SNR thresholds, namely $\gamma = 0 \hspace{0.1cm}\text{dB}$ and $\gamma = 10\hspace{0.1cm} \text{dB}$, corresponding to $p_{\text{s}}=0.922$ and $p_{\text{s}}=0.445$, respectively.
\par The time-averaged reconstruction error under the semantics-aware, change-aware, uniform, and RS policies for a three-state DTMC and the BDMP information source models are shown in Table \ref{Table1} and \ref{Table2} for different values of $p$, $q$, and $p_{s}$. In the uniform policy, a sample is acquired every $5$ time slots. A first observation is that the semantics-aware policy outperforms all other sampling policies, especially when the source is rapidly changing.
\begin{table}[]
	\centering
	\caption{\scriptsize{Time-averaged reconstruction error under DTMC model for $p_{\alpha^{\text{s}}}=0.7$, and different values of , $p_{\text{s}}$, $p$ and $q = 1-2p$.}}
	\label{Table1}
	\begin{tabular}{|c|c|c|c|c|c|c|}
		\hline
		${p}$ &${q}$& ${p_{\text{s}}}$ & $\text{{Semantics-aware }}$& ${\text{Change-aware}}$&${\text{Uniform}}$&${\text{RS}}$\\
		\hline
		{0.1}  & {0.8}&{0.922}  & {0.016} & {0.075}&{0.322}&{0.094}\\ \hline
		{0.1} & {0.8}&{0.445}  & {0.181}  & {0.434} &{0.485}&{0.266} \\ \hline
	{0.3}  & {0.4}&{0.922}  & {0.047} & {0.075}&{0.529}&{0.220}\\ \hline
		{0.3} & {0.4}&{0.445}  & {0.352}  & {0.434} &{0.601}&{0.443} \\ \hline
	\end{tabular}
\end{table}
\begin{table}[]
	\centering
	\caption{\scriptsize{Time-averaged reconstruction error under BDMP model for $p_{\alpha^{\text{s}}}=0.7$, and different values of $p_{\text{s}}$, $p$ and $q$.}}
	\label{Table2}
	\begin{tabular}{|c|c|c|c|c|c|c|}
		\hline
		${p}$ &$q$& ${p_{\text{s}}}$ & $\text{{Semantics-aware }}$& ${\text{Change-aware}}$&${\text{Uniform}}$&${\text{RS}}$\\
		\hline
		{0.1}  & {0.2}& {0.922}  & {0.014} & {0.074}&{0.259}&{0.078}\\ \hline
		{0.1} & {0.2}&{0.445}  & {0.148}  & {0.413} &{0.394}&{0.215} \\ \hline
	{0.2}  & {0.7}&{0.922}  & {0.029} & {0.073}&{0.315}&{0.133}\\ \hline
		{0.2} &{0.7}& {0.445}  & {0.211}  & {0.398} &{0.367}&{0.266} \\ \hline
	\end{tabular}
\end{table}
\par
\begin{figure}[!t]
	\centering
 \subfigure[DTMC model]{\includegraphics[width=0.485\linewidth, clip]{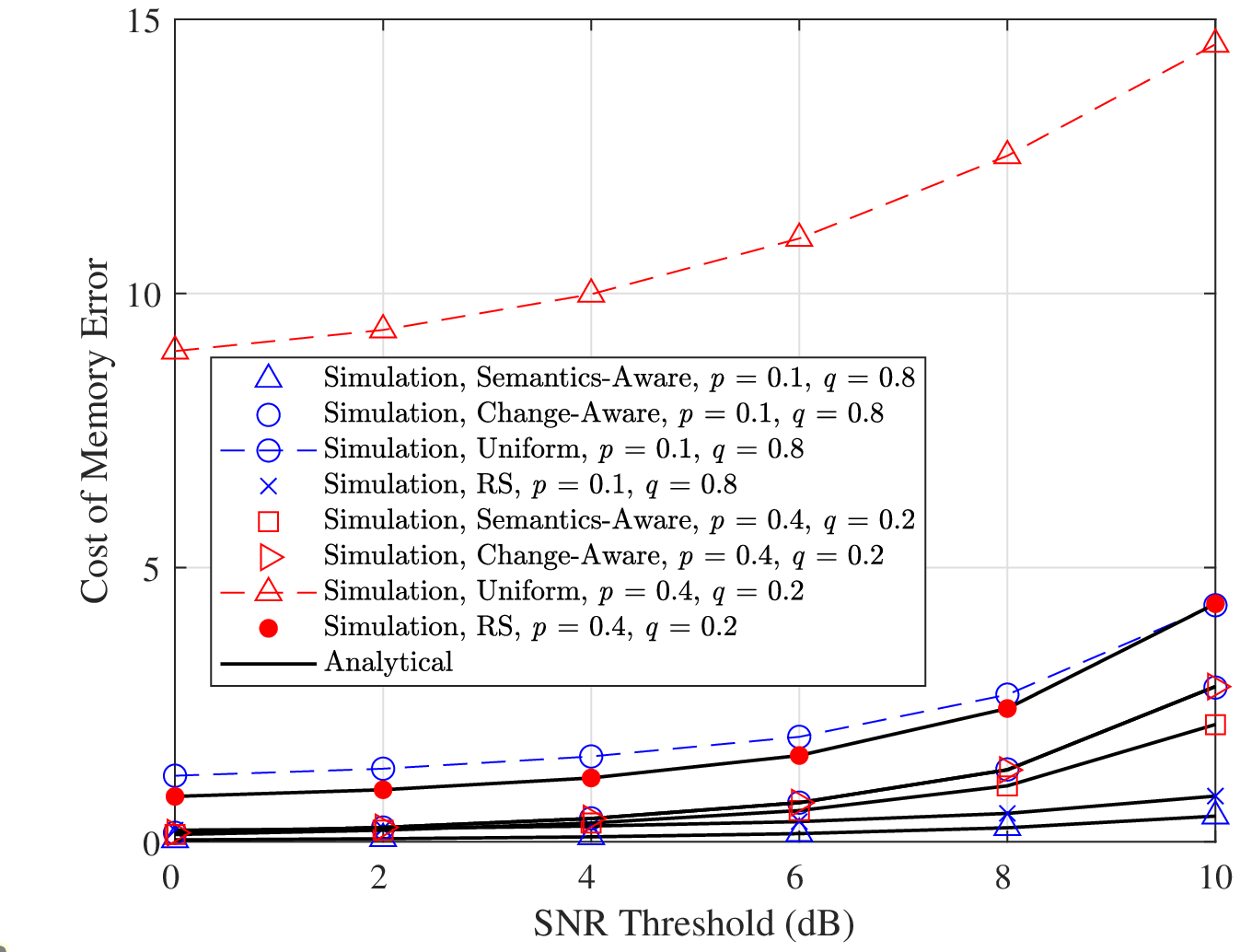}
\label{Penalization_MemoryError_M1}}
 \subfigure[BDMP model]{\centering
 	\includegraphics[width=0.485\linewidth, clip]{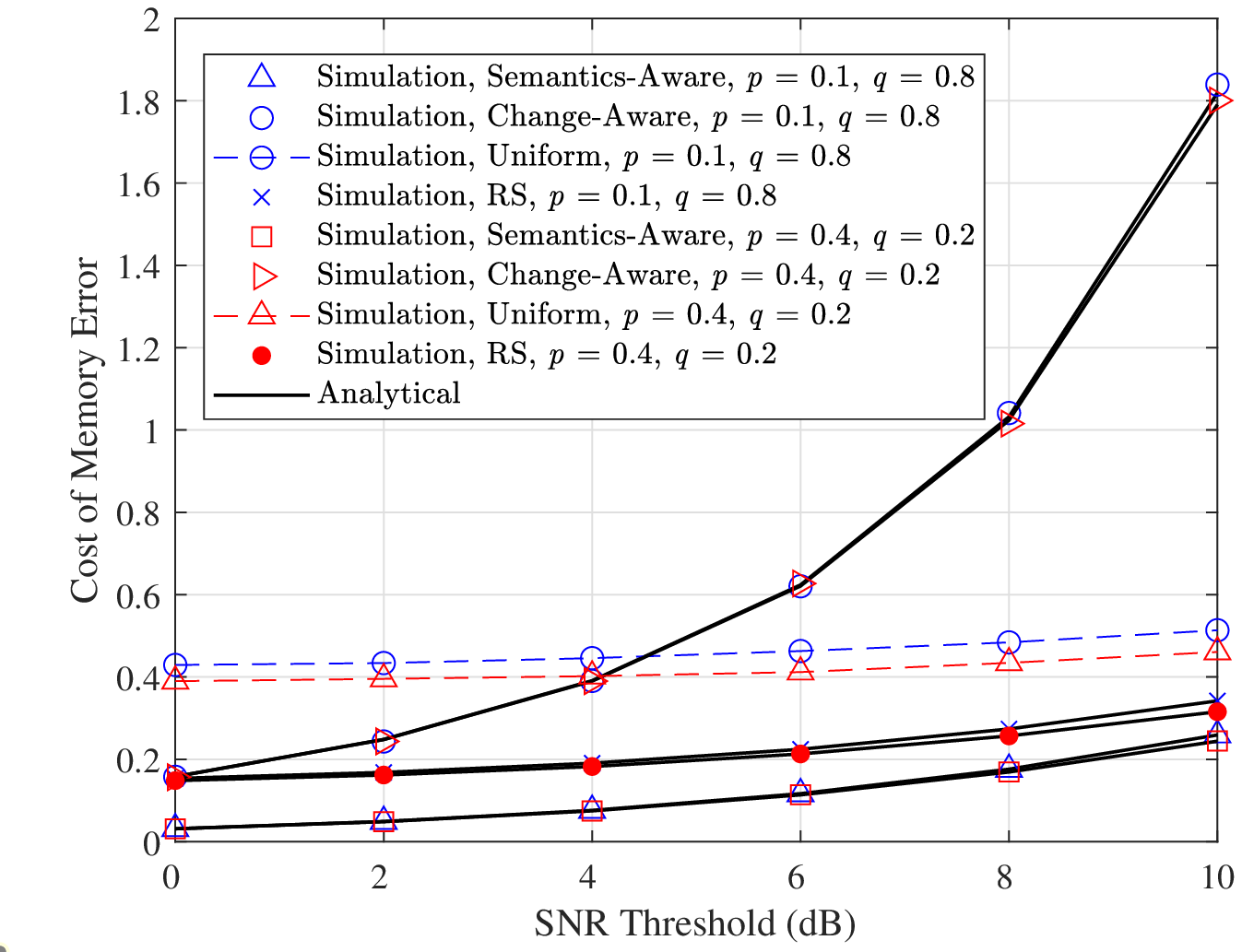}
	\label{Penalization_MemoryError_M2}}
 \caption{Cost of memory error using the DTMC and BDMP models as a function of $\gamma$, for $n = 10$, $\kappa = 2$, $p_{\alpha^{\text{s}}} = 0.7$, and different values of $p$ and $q$.}
 \end{figure}

 Figs. \ref{Penalization_MemoryError_M1} and \ref{Penalization_MemoryError_M2} show the cost of memory error for the DTMC and the BDMP model, respectively, as a function of $\gamma$ for $n = 10$, $\kappa = 2$, $p_{\alpha^{\text{s}}} = 0.7$,  and different values of $p$ and $q$. As seen in these figures, as $\gamma$ increases, the cost of memory error increases. This is because when $\gamma$ increases, the success probability $p_{\text{s}}$ decreases, thus a transmitted sample is successfully decoded with a lower probability. Note also that the semantics-aware policy exhibits smaller cost of memory error compared to all other policies. Thus, the semantics-aware policy does not allow the system to operate in an erroneous state for several consecutive time slots.
\par
\begin{figure}[!t]
	\centering
	\includegraphics[width=0.6\linewidth, clip]{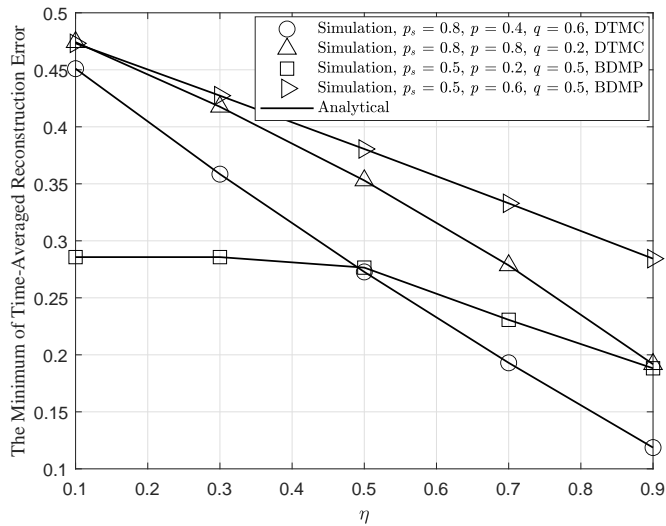}
	\caption{Minimum time-averaged reconstruction error under RSC policy for the DTMC and BDMP models.}
	\label{OptimizationProblem1_M1M2}
\end{figure}

 The performance of the optimal RSC policy in terms of time-averaged  reconstruction error as a function of $p_{\text{s}}$, $p$, and $q$ is illustrated in Fig. \ref{OptimizationProblem1_M1M2}. As seen in that figure and Tables \ref{Table1_RSC_M1}, \ref{Table2_RSC_M1}, \ref{Table3_RSC_M2} and \ref{Table4_RSC_M2}, the time-averaged reconstruction error has a smaller minimum value when $\eta$ increases. This is because $\eta$ is the threshold of the sampling probability, $p_{\alpha^{\text{s}}}\leqslant \eta$, thus a higher value of $\eta$ results in higher sampling probability which in turn decreases the time-averaged reconstruction error. Another interesting observation is that by modeling the information source as a Birth-Death process, when $\displaystyle\eta<\frac{q-p}{p_{\text{s}}(1+q-p)}$ (in this simulation it is assumed that the receiver at $t = 0$ is at the state $0$) the optimal sampling policy is $P^{*}_{E} = P^{\text{NS}}_{E}$, i.e., not performing any sampling. 
\par In Tables \ref{Table_Optimal_M1} and \ref{Table_Optimal_M2}, we consider DTMC and BDMP models, respectively, and show the minimum time-averaged reconstruction error under a sampling cost constraint for $p_{\text{s}}=0.5$, $\eta=0.5$ and different values of $p$ and $q$. We observe that the semantics-aware policy outperforms the optimal RSC policy for $p\leqslant\frac{\eta p_{\text{s}}}{1-2\eta+2\eta p_{\text{s}}}$ (DTMC model) and $\frac{2pq}{(p+q)\big[p(1-p_{\text{s}})+q+p_{\text{s}}(1-q)\big]}\leqslant \eta$ (BDMP model) when the source is slowly varying (see Appendix \ref{Appendix_AvgCost}). Note that the optimal values with red color for the semantics-aware, change-aware, and RS policies are obtained for values of $p$, $q$ and $p_{\alpha^{\text{s}}}$ that violate the constraint requirement. This means that in an unconstrained scenario, the performance of the optimal RS and the semantics-aware policies is the same, however in that case, the optimal solution for the RS policy turns out to sample and transmit on every time slot; this has the undesirable result of generating an excessive amount of samples. In sharp contrast, a key result of this work is that when a constraint on the sampling cost is imposed, which is practically relevant scenario, the proposed RS policy outperforms all other sampling policies for a rapidly changing source.
\begin{table}[]
\centering
\caption{\scriptsize{Minimum time-averaged reconstruction error under RSC policy using  DTMC model for $p_{\text{s}} = 0.8$, $p = 0.4$, and $q = 0.6$.}}
\label{Table1_RSC_M1}
\begin{tabular}{|c|c|c|c|c|}
\hline
$\eta$ & $p_{\alpha}^{*}$ & $P^{\text{NS}}_{E}$& Minimum time-averaged reconstruction error \\ \hline
0.1  & 0.1  & 0.5 & 0.4510\\ \hline
0.3 & 0.3  & 0.5  & 0.3585                    \\ \hline
0.5 & 0.5 & 0.5 & 0.2727\\ \hline
0.7 & 0.7 & 0.5 & 0.1930\\ \hline
0.9 & 0.9 & 0.5 & 0.1186\\ \hline
\end{tabular}
\end{table}
\begin{table}[]
\centering
\caption{\scriptsize{Minimum time-averaged reconstruction error under RSC policy using  DTMC model for $p_{\text{s}}= 0.8$, $p = 0.8$, and $q = 0.2$.}}
\label{Table2_RSC_M1}
\begin{tabular}{|c|c|c|c|c|}
\hline
$\eta$ & $p_{\alpha}^{*}$ & $P^{\text{NS}}_{E}$& Minimum time-averaged reconstruction error \\ \hline
0.1  & 0.1  & 0.5 & 0.4742\\ \hline
0.3 & 0.3  & 0.5  & 0.4176                   \\ \hline
0.5 & 0.5 & 0.5 & 0.3529\\ \hline
0.7 & 0.7 & 0.5 & 0.2785\\ \hline
0.9 & 0.9 & 0.5 & 0.1918\\ \hline
\end{tabular}
\end{table}

\begin{table}[]
\centering
\caption{\scriptsize{Minimum time-averaged reconstruction error under RSC policy using  BDMP model for $p_{\text{s}}= 0.5$, $p = 0.2$, and $q = 0.5$.}}
\label{Table3_RSC_M2}
\begin{tabular}{|c|c|c|c|c|}
\hline
$\eta$ & $p_{\alpha}^{*}$ & $P^{\text{NS}}_{E}$& Minimum time-averaged reconstruction error \\ \hline
0.1  & 0  & 0.2857 & 0.2857\\ \hline
0.3 & 0  & 0.2857  & 0.2857                   \\ \hline
0.5 & 0.5 & 0.2857 & 0.2765\\ \hline
0.7 & 0.7 & 0.2857 & 0.2307\\ \hline
0.9 & 0.9 & 0.2857 & 0.1882\\ \hline
\end{tabular}
\end{table}
\begin{table}[]
\centering
\caption{\scriptsize{Minimum time-averaged reconstruction error under RSC policy using  BDMP model for $p_{\text{s}}= 0.5$, $p = 0.6$, and $q = 0.5$.}}
\label{Table4_RSC_M2}
\begin{tabular}{|c|c|c|c|c|}
\hline
$\eta$ & $p_{\alpha}^{*}$ & $P^{\text{NS}}_{E}$& Minimum time-averaged reconstruction error \\ \hline
0.1  & 0.1  & 0.5455 & 0.4732\\ \hline
0.3 & 0.3  & 0.5455  & 0.4273                   \\ \hline
0.5 & 0.5 & 0.5455 & 0.3805\\ \hline
0.7 & 0.7 & 0.5455 & 0.3329\\ \hline
0.9 & 0.9 & 0.5455 & 0.2844\\ \hline
\end{tabular}
\end{table}

\begin{table}[]
	\centering
	\caption{\scriptsize{Minimum time-averaged reconstruction error using the DTMC model for $p_{\text{s}}=0.5$, $\eta=0.5$.}}
	\label{Table_Optimal_M1}
	\begin{tabular}{|c|c|c|c|c|c|c|}
		\hline
		$p$ &$q$& $\text{Semantics-aware}$& $\text{Change-aware}$&$\text{Uniform}$&$\text{RSC}$&$\textcolor{red}{\text{RS}}$\\
		\hline
		0.1  &0.9& 0.083 & 0.333&0.299&0.187&\textcolor{red}{0.083}\\ \hline
		0.3 &0.7& 0.187  & 0.333 &0.417&0.321 &\textcolor{red}{0.187}\\ \hline
		0.5   &0.5& 0.250 & 0.333&0.450&0.375&\textcolor{red}{0.250}\\ \hline
		0.7  &0.3& \textcolor{red}{0.291}  & \textcolor{red}{0.333}&0.464&0.404&\textcolor{red}{0.291}\\ \hline
		0.9  &0.1& \textcolor{red}{0.321}  & \textcolor{red}{0.333}&0.468&0.422&\textcolor{red}{0.321} \\ \hline
	\end{tabular}
\end{table}
\begin{table}[]
	\centering
	\caption{\scriptsize{Minimum time-averaged reconstruction error using the BDMP model for $p_{\text{s}}=0.5$, $\eta=0.5$.}}
	\label{Table_Optimal_M2}
	\begin{tabular}{|c|c|c|c|c|c|c|}
		\hline
		$p$ &$q$& $\text{Semantics-aware}$& $\text{Change-aware}$&$\text{Uniform}$&$\text{RSC}$&$\textcolor{red}{\text{RS}}$\\
		\hline
		0.1  &0.2& 0.102 & 0.333&0.311&0.210&\textcolor{red}{0.102}\\ \hline
		0.3 &0.4& 0.201  & 0.333 &0.421&0.332 &\textcolor{red}{0.201}\\ \hline
		0.5   &0.6& \textcolor{red}{0.259} & \textcolor{red}{0.333}&0.451&0.380&\textcolor{red}{0.259}\\ \hline
		0.7  &0.8& \textcolor{red}{0.298}  & \textcolor{red}{0.333}&0.464&0.407&\textcolor{red}{0.298}\\ \hline
		0.9  &0.95& \textcolor{red}{0.324}  & \textcolor{red}{0.333}&0.468&0.423&\textcolor{red}{0.324} \\ \hline
	\end{tabular}
\end{table}
\begin{table}[]
\centering
\caption{\scriptsize{Optimal value of $n$ and average consecutive error using BDMP model as a function of $p$ for $q = 0.2$, $p_{\text{s}}=0.5$, and $\eta = 0.2$.}}
\label{Table1_optimizationproblem2}
\begin{tabular}{|c|c|c|c|c|}
\hline
$p$ & $P^{\text{NS}}_{E}$  & $P^{\text{AS}}_{E}$& $n^{*}$& $\bar{C}_{E}(n^{*})$ \\ \hline
$0.1$ & $0.3333$ & $0.1026$& $2$& $0.4084$ \\ \hline
$0.3$ & $0.6$ & $0.16$& $3$& $0.9654$\\ \hline
$0.5$& $0.7143$  & $0.1681$  & $3$ & $1.1783$\\ \hline
$0.7$ & $0.7778$ & $0.1637$  & $3$ & $1.2853$\\  \hline
$0.9$ & $0.8182$ & $0.1558$  & $4$ & $1.6885$\\  \hline
\end{tabular}
\end{table}

\begin{table}[]
\centering
\caption{\scriptsize{Optimal value of $n$ and average consecutive error using BDMP model as a function of $p$ for $q = 0.2$, $p_{\text{s}}=0.5$, and $\eta = 0.6$.}}
\label{Table2_optimizationproblem2}
\begin{tabular}{|c|c|c|c|c|}
\hline
$p$ & $P^{\text{NS}}_{E}$  & $P^{\text{AS}}_{E}$& $n^{*}$& $\bar{C}_{E}(n^{*})$ \\ \hline
$0.1$ & $0.3333$ & $0.1026$& $1$& $0.3018$ \\ \hline
$0.3$ & $0.6$ & $0.16$& $1$& $0.4960$\\ \hline
$0.5$& $0.7143$  & $0.1681$  & $1$ & $0.5553$\\ \hline
$0.7$ & $0.7778$ & $0.1637$  & $1$ & $0.5762$\\  \hline
$0.9$ & $0.8182$ & $0.1558$  & $1$ & $0.5831$\\  \hline
\end{tabular}
\end{table}
\par Tables \ref{Table1_optimizationproblem2} and \ref{Table2_optimizationproblem2} show the optimal value of $n$ and the average of consecutive error using the BDMP model as a function of $p$, for $q = 0.2$, $p_{\text{s}}=0.5$, and selected values of $\eta$. We observe that when $p$ increases, the optimal average consecutive error increases (it is assumed without loss of generality that $X_0 = 0$) . This is because as $p$ increases, $P^{\text{NS}}_{E}$ increases and therefore sampling should start at a higher value of $n$ to satisfy the time-averaged sampling cost, and this eventually increases the optimal average consecutive error. Moreover, note that when $\eta$ increases, we can sample and transmit at a much lower value of $n$ without violating the sampling cost constraint. For example, for $p=0.1, 0.3, 0.5, 0.7, 0.9$, $q = 0.2$, $p_{\text{s}}=0.5$, and $\eta = 0.6$, the optimal value of $n$ is equal to $1$. This means that the sampler can always perform sampling without violating the sampling cost constraint.
\section{Conclusions}
\label{Conclusions}
\par We investigated a time-slotted communication system where a device performs joint sampling and transmission over a wireless channel to track the evolution of a Markov source. Two models are considered for the information source: an $N$-state Markov chain and an $N$-state Birth-Death Markov process. We considered four sampling and transmission policies, namely uniform, change-aware, semantics-aware, and randomized stationary policies. Then, we derived general expressions for the transition probabilities under randomized stationary policy and studied the performance of the system in terms of the time-averaged reconstruction error and its variance, the cost of actuation error, the average consecutive error, and the cost of memory error for all aforementioned policies. Furthermore, we formulated two optimization problems. In the first one, the goal is to find the optimal randomized stationary policy that minimizes the time-averaged reconstruction error under a sampling cost constraint. Its solution shows that the semantics-aware policy performs the best except when the source is rapidly evolving and there is a sampling cost constraint, in which cases the randomized stationary policy is superior. In the second optimization problem, we propose a \emph{wait-then-generate} sampling policy to minimize the average consecutive error while keeping the time-averaged sampling cost constraint less than a given threshold. Capitalizing on the solution of this optimization problem, we derive a threshold that can be utilized by the sampler so as to perform sampling that satisfies the constraint and minimizes the objective function.

\appendix
\subsection{Proof of Lemma {\ref{lemma_P00}}}
\label{Appendixlemma1}
Using eq. \eqref{trans_prob2},  \eqref{c1_c2}, and \eqref{Cond_Prob_X_E}, we can derive $P_{0,0}$ as
\begin{align}
	\label{P00}
	P_{0,0} &=\sum_{n=0}^{N-1} \frac{1}{\mathrm{Pr}\big[E_{t}=0\big]}\mathrm{Pr}\big[E_{t+1} = 0\big|X_{t} = n, \hat{X}_{t} = n\big]\mathrm{Pr}\big[X_{t} = n, \hat{X}_{t} = n\big]\notag\\
	&=\sum_{n=0}^{N-1} \frac{1}{N}\mathrm{Pr}\big[E_{t+1} = 0\big|X_{t} = n, \hat{X}_{t} = n\big].
\end{align}

We now calculate the conditional probability in \eqref{P00}. To this end, we first note that the receiver has perfect knowledge of the process $X_{t}$ at time slot $t$. Therefore, the system will be in synced state at time slot $t+1$, i.e., $E_{t+1} \!=\! 0$, if the state of the process $X_{t}$ does not change, which happens with probability $q$. In addition, $E_{t+1}\!=\!0$ when the state of the source changes to one of the remaining $N\!-\!1$ states and the sample is successfully decoded by the receiver. This event occurs with probability $(N-1)p{H}_{1}$. Then \eqref{P00} can be written as 
\begin{align}
	\label{P00_2}
	P_{0,0} =  q+(N-1)p{H}_{1}.
\end{align}
\subsection{Proof of Lemma {\ref{lemma_PI_ij_DTMCmodel}}}
\label{Appendixlemma2}
\begin{figure}[!t]
	\centering
  \scriptsize
 \subfigure[DTMC model]{
 \begin{tikzpicture}[start chain=going left,->,>=latex,node distance=2.5cm]

		\node[state]    (A)                     {$(0,0)$};
		\node[state]    (B)[above right of=A]   {$(0,1)$};
		\node[state]    (C)[below right of=A]   {$(1,0)$};
		\node[state]    (D)[below right of=B]   {$(1,1)$};
		\path
		(A) edge[loop left]     node{$1-p$}  (A)

		edge[bend left=15,above]  node{$pp_{\alpha^{\text{s}}}p_{\text{s}}$}  (D)
		edge[bend right=15,left]    node{$p(1-p_{\alpha^{\text{s}}}p_{\text{s}})$}  (C)
		(B) edge[loop above]  node{$(1-p)(1-p_{\alpha^{\text{s}}}p_{\text{s}})$}     (B)
		edge[left=15] node{$(1-p)p_{\alpha^{\text{s}}}p_{\text{s}}$}    (A)
		edge[bend right=15,left] node{$p$}    (D)
		
		(C) edge[loop below]  node{$(1-p)(1-p_{\alpha^{\text{s}}}p_{\text{s}})$}     (C)
		edge[bend right=15,right] node{$p$}    (A)
		
		edge[ left=15,right] node{$(1-p)p_{\alpha^{\text{s}}}p_{\text{s}}$}    (D)
		(D) edge[loop right]    node{$1-p$}     (D)
		edge[bend right=15,right]  node{$p(1-p_{\alpha^{\text{s}}}p_{\text{s}})$}  (B)
		
		edge[bend left=15,above]     node{$pp_{\alpha^{\text{s}}}p_{\text{s}}$}         (A);
	\end{tikzpicture}
 \label{2DimMarkovChainM1}
 }
 \hfill
 \subfigure[BDMP model]{
 \begin{tikzpicture}[start chain=going left,->,>=latex,node distance=2.5cm]
		
		\node[state]    (A)                     {$(0,0)$};
		\node[state]    (B)[above right of=A]   {$(0,1)$};
		\node[state]    (C)[below right of=A]   {$(1,0)$};
		\node[state]    (D)[below right of=B]   {$(1,1)$};
		\path
		(A) edge[loop left]     node{$1-p$}  (A)
		
		edge[bend left=15,above]  node{$pp_{\alpha^{\text{s}}}p_{\text{s}}$}  (D)
		edge[bend right=15,left]    node{$p(1-p_{\alpha^{\text{s}}}p_{\text{s}})$}  (C)
		(B) edge[loop above]  node{$(1-p)(1-p_{\alpha^{\text{s}}}p_{\text{s}})$}     (B)
		edge[left=15] node{$(1-p)p_{\alpha^{\text{s}}}p_{\text{s}}$}    (A)
		edge[bend right=15,left] node{$p$}    (D)
		
		(C) edge[loop below]  node{$(1-q)(1-p_{\alpha^{\text{s}}}p_{\text{s}})$}     (C)
		edge[bend right=15,right] node{$q$}    (A)
		
		edge[ left=15,right] node{$(1-q)p_{\alpha^{\text{s}}}p_{\text{s}}$}    (D)
		(D) edge[loop right]    node{$1-q$}     (D)
		edge[bend right=15,right]  node{$q(1-p_{\alpha^{\text{s}}}p_{\text{s}})$}  (B)
		
		edge[bend left=15,above]     node{$qp_{\alpha^{\text{s}}}p_{\text{s}}$}         (A);
	\end{tikzpicture}
 \label{2DimMarkovChainM2}
 }
	\vspace*{1ex}
	\caption{Two-dimensional Markov chain describing the joint status of the information source regarding the current state at the original source for a two-state model for the information source.}
 \label{2DimMarkovChainM1_M2}
\end{figure}
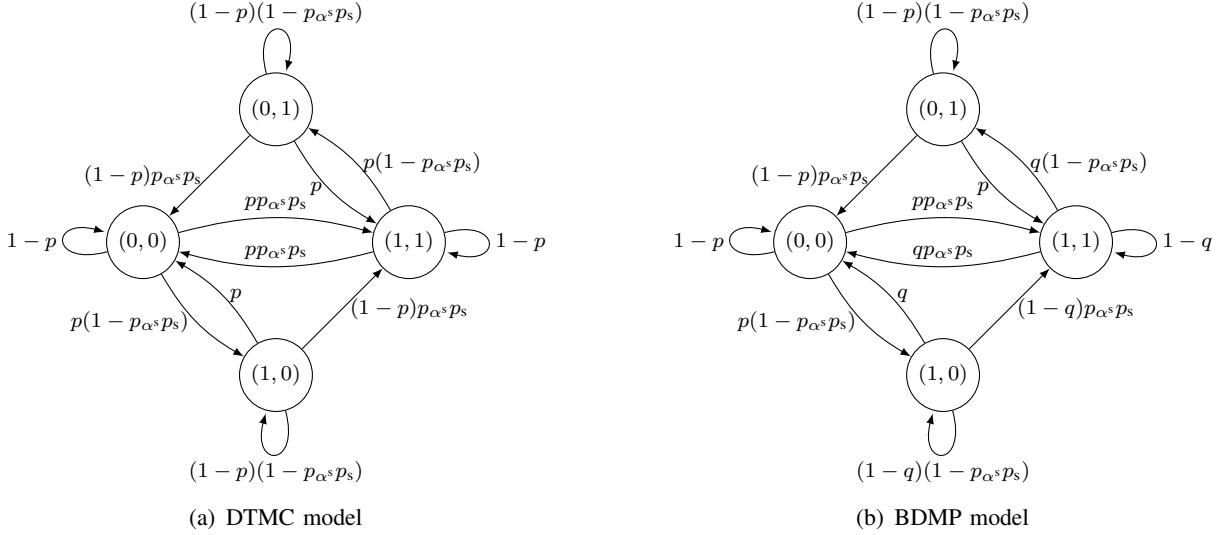
\par To obtain $\pi_{i,j}$, we depict the two-dimensional DTMC describing the joint status of the information source regarding the current state at the original source, i.e., $(X_{t}, \hat{X}_{t})$ as Fig. \ref{2DimMarkovChainM1}. Now, using Fig. \ref{2DimMarkovChainM1}, we can obtain the state stationary $\pi_{i,j}$ for the randomized stationary policy as

\begin{align}
	\label{pij_M1_N2}
	\pi_{0,0}&=\frac{p+(1-p)p_{\alpha^{\text{s}}}p_{\text{s}}}{4p+2p_{\alpha^{\text{s}}}p_{\text{s}}-4pp_{\alpha^{\text{s}}}p_{\text{s}}},\hspace{0.2cm}
	\pi_{0,1}=\frac{p(1-p_{\alpha^{\text{s}}}p_{\text{s}})}{4p+2p_{\alpha^{\text{s}}}p_{\text{s}}-4pp_{\alpha^{\text{s}}}p_{\text{s}}}\notag\\
	\pi_{1,0}&=\frac{p(1-p_{\alpha^{\text{s}}}p_{\text{s}})}{4p+2p_{\alpha^{\text{s}}}p_{\text{s}}-4pp_{\alpha^{\text{s}}}p_{\text{s}}},\hspace{0.2cm}
	\pi_{1,1}=\frac{p+(1-p)p_{\alpha^{\text{s}}}p_{\text{s}}}{4p+2p_{\alpha^{\text{s}}}p_{\text{s}}-4pp_{\alpha^{\text{s}}}p_{\text{s}}}.
\end{align}
For the change-aware policy, eq. \eqref{pij_M1_N2} can be written as
\begin{align}
	\label{CA_M1_N2}
	\pi_{0,0}  =\pi_{1,1}= \frac{1}{4-2p_{\text{s}}},\hspace{0.2cm}
	\pi_{0,1} =\pi_{1,0}= \frac{1-p_{\text{s}}}{4-2p_{\text{s}}}.
\end{align}
Furthermore, for the semantics-aware policy $\pi_{i,j}$ is given by
\begin{align}
	\label{SA_M1_N2}
	\pi_{0,0}=\pi_{1,1} = \frac{p+p_{\text{s}}-pp_{\text{s}}}{4p+2p_{\text{s}}-4pp_{\text{s}}},\hspace{0.2cm}
	\pi_{0,1}=\pi_{1,0} = \frac{p(1-p_{\text{s}})}{4p+2p_{\text{s}}-4pp_{\text{s}}}.
\end{align}
Using a similar procedure, the state stationary $\pi_{i,j}$ for the randomized stationary policy and a three-state DTMC describing the evolution of the information source can be obtained as
\begin{align}
	\label{pij_M1_N3}
	\pi_{i,i} &= \frac{p+p_{\alpha^{\text{s}}}p_{\text{s}}-pp_{\alpha^{\text{s}}}p_{\text{s}}}{9p+3p_{\alpha^{\text{s}}}p_{\text{s}}-9pp_{\alpha^{\text{s}}}p_{\text{s}}}, \hspace{0.2cm} \forall i \in \{0, 1, 2\}\notag\\
	\pi_{i,j} &= \frac{p-p p_{\alpha^{\text{s}}}p_{\text{s}}}{9p+3p_{\alpha^{\text{s}}}p_{\text{s}}-9pp_{\alpha^{\text{s}}}p_{\text{s}}}, \hspace{0.2cm} \forall i,j \in \{0, 1, 2\}, i\neq j.
\end{align}
For the change-aware policy, $\pi_{i,j}$ in eq. \eqref{pij_M1_N3}  can be written as
\begin{align}
	\label{CA_M1_N3}
	\pi_{i,i}=\frac{1+p_{\text{s}}}{9-2p_{\text{s}}},\hspace{0.1cm} \forall i \in \{0, 1, 2\},\hspace{0.2cm}
	\pi_{i,j}= \frac{1-p_{\text{s}}}{9-3p_{\text{s}}},\hspace{0.1cm}\forall i,j \in \{0, 1, 2\}, i\neq j.
\end{align}
For the semantics-aware policy, $\pi_{i,j}$  can be written as 
\begin{align}
	\label{SA_M1_N3_2}
	\pi_{i,i}= \frac{p+p_{\text{s}}-pp_{\text{s}}}{9p+3p_{\text{s}}-9pp_{\text{s}}}, \hspace{0.1cm} \forall i \in \{0, 1, 2\},\hspace{0.2cm}
	\pi_{i,j}= \frac{p(1-p_{\text{s}})}{9p+3p_{\text{s}}-9pp_{\text{s}}}, \hspace{0.1cm} \forall i,j \in \{0, 1, 2\}, i\neq j.
\end{align}
\subsection{Proof of Lemma {\ref{lemma_PI_ij_BirthDeathModel}}}
\label{Appendixlemma3}
\par In order to calculate $\pi_{i,j}$, we depict a two-dimensional DTMC as Fig. \ref{2DimMarkovChainM2}. Now, using Fig. \ref{2DimMarkovChainM2}, the state stationary $\pi_{i,j}$ for the randomized stationary policy can be obtained as

\begin{align}
	\label{pij_M2_N2_2}
	\pi_{0,0} &=\frac{q\big[q+(1-q)p_{\alpha^{\text{s}}}p_{\text{s}}\big]}{(p+q)\big[p(1-p_{\alpha^{\text{s}}}p_{\text{s}})\!+\!q\!+\!(1-q)p_{\alpha^{\text{s}}}p_{\text{s}}\big]},\hspace{0.1cm}
	\pi_{0,1} =\frac{pq(1-p_{\alpha^{\text{s}}}p_{\text{s}})}{(p+q)\big[p(1-p_{\alpha^{\text{s}}}p_{\text{s}})\!+\!q\!+\!(1-q)p_{\alpha^{\text{s}}}p_{\text{s}}\big]}\notag\\
	\pi_{1,0} &=\frac{pq(1-p_{\alpha^{\text{s}}}p_{\text{s}})}{(p+q)\big[p(1-p_{\alpha^{\text{s}}}p_{\text{s}})\!+\!q\!+\!(1-q)p_{\alpha^{\text{s}}}p_{\text{s}}\big]},\hspace{0.1cm}
	\pi_{1,1} =\frac{p\big[p+(1-p)p_{\alpha^{\text{s}}}p_{\text{s}}\big]}{(p+q)\big[p(1-p_{\alpha^{\text{s}}}p_{\text{s}})\!+\!q\!+\!(1-q)p_{\alpha^{\text{s}}}p_{\text{s}}\big]}.
\end{align}
For the change-aware policy, $\pi_{i,j}$ can be written as
\begin{align}
	\label{pij_M2_CA_N2}
	\pi_{0,0}&=\frac{q}{(p+q)(2-p_{\text{s}})},\hspace{0.1cm}
	\pi_{0,1}=\frac{q(1-p_{\text{s}})}{(p+q)(2-p_{\text{s}})}\notag\\
	\pi_{1,0}&=\frac{p(1-p_{\text{s}})}{(p+q)(2-p_{\text{s}})},\hspace{0.1cm}
	\pi_{1,1}=\frac{p}{(p+q)(2-p_{\text{s}})}.
\end{align}
Moreover, for the semantics-aware policy, we can calculate $\pi_{i,j}$ as
\begin{align}
	\label{pij_M2_SA_N2}
	\pi_{0,0} &=\frac{q\big[q+(1-q)p_{\text{s}}\big]}{(p+q)\big[p(1-p_{\text{s}})+q+(1-q)p_{\text{s}}\big]},\hspace{0.1cm}
	\pi_{0,1} =\frac{pq(1-p_{\text{s}})}{(p+q)\big[p(1-p_{\text{s}})+q+(1-q)p_{\text{s}}\big]}\notag\\
	\pi_{1,0} &=\frac{pq(1-p_{\text{s}})}{(p+q)\big[p(1-p_{\text{s}})+q+(1-q)p_{\text{s}}\big]},\hspace{0.1cm}
	\pi_{1,1} =\frac{p\big[p+(1-p)p_{\text{s}}\big]}{(p+q)\big[p(1-p_{\text{s}})+q+(1-q)p_{\text{s}}\big]}.
\end{align}
Similarly to the two-state case, the state stationary $\pi_{i,j}$ for the randomized stationary policy and  a three-state BDMP model describing the evolution of the information source can be calculated as
    \begin{align}
	\label{pij_M2_N3}
	\pi_{0,0}&=\frac{q^{2}\big[pp_{\alpha^{\text{s}}}p_{\text{s}}(1-p_{\alpha^{\text{s}}}p_{\text{s}})+\big(q+(1-q)p_{\alpha^{\text{s}}}p_{\text{s}}\big)\big]}{I(p,q,p_{\alpha^{\text{s}}},p_{\text{s}})}\notag\\
	\pi_{0,1} &=\pi_{1,0}= \frac{pq^{2}\big[1-p_{\alpha^{\text{s}}}p_{\text{s}}\big]\big[q+(1-q)p_{\alpha^{\text{s}}}p_{\text{s}}\big]}{I(p,q,p_{\alpha^{\text{s}}},p_{\text{s}})}\notag\\
	\pi_{1,1}&=\frac{pq\big[p_{\alpha^{\text{s}}}p_{\text{s}}+p(1-p_{\alpha^{\text{s}}}p_{\text{s}})\big]\big[q+(1-q)p_{\alpha^{\text{s}}}p_{\text{s}}\big]}{I(p,q,p_{\alpha^{\text{s}}},p_{\text{s}})}\notag\\
	\pi_{0,2}&=\pi_{2,0}=\frac{p^{2}q^{2}(1-p_{\alpha^{\text{s}}}p_{\text{s}})^{2}}{I(p,q,p_{\alpha^{\text{s}}},p_{\text{s}})}\notag\\
	\pi_{1,2}&=\pi_{2,1}=\frac{p^{2}q\big[1-p_{\alpha^{\text{s}}}p_{\text{s}}\big]\big[p+(1-p)p_{\alpha^{\text{s}}}p_{\text{s}}\big]}{I(p,q,p_{\alpha^{\text{s}}},p_{\text{s}})}\notag\\
	\pi_{2,2}&=\frac{p^{2}\big[2pp_{\alpha^{\text{s}}}p_{\text{s}}(1-p_{\alpha^{\text{s}}}p_{\text{s}})+p^{2}(1-p_{\alpha^{\text{s}}}p_{\text{s}})^{2}+p_{\alpha^{\text{s}}}p_{\text{s}}\big(q+(1-q)p_{\alpha^{\text{s}}}p_{\text{s}}\big)\big]}{I(p,q,p_{\alpha^{\text{s}}},p_{\text{s}})},
\end{align}
where $I(p,q,p_{\alpha^{\text{s}}},p_{\text{s}})$ can be written as
\begin{align}
	&I(p,q,p_{\alpha^{\text{s}}},p_{\text{s}})\notag\\
	&= \left(p^2+pq+q^2\right)\Big[p\big(1-p_{\alpha^{\text{s}}}p_{\text{s}}\big)\big(q+(2-q)p_{\alpha^{\text{s}}}p_{\text{s}}\big)+\big(q+(1-q)p_{\alpha^{\text{s}}}p_{\text{s}}\big)^{2}+p^{2}\big(1-p_{\alpha^{\text{s}}}p_{\text{s}}\big)^{2}\Big].
\end{align}
For the change-aware policy $\pi_{i,j}$ can be obtained as
\begin{align}
	\label{pij_M2_CA_N3}
	\pi_{0,0}&=\frac{q^{2}\big[q+pp_{\text{s}}\big(2-p_{\text{s}}\big)\big]}{\big(p+q\big)\big(2-p_{\text{s}}\big)\big(p^{2}+pq+q^{2}\big)},\hspace{0.1cm}
	\pi_{0,1}=\frac{q^{2}\big(1-p_{\text{s}}\big)}{\big(2-p_{\text{s}}\big)\big(p^2+pq+q^2\big)}\notag\\
	\pi_{0,2}&=\frac{pq^{2}\big(1-p_{\text{s}}\big)^{2}}{\big(p+q\big)\big(2-p_{\text{s}}\big)\big(p^2+pq+q^2\big)},\hspace{0.1cm}
	\pi_{1,0}=\frac{pq^{2}\big(1-p_{\text{s}}\big)}{\big(p+q\big)\big(2-p_{\text{s}}\big)\big(p^2+pq+q^2\big)}\notag\\
	\pi_{1,1}&=\frac{pq}{\big(2-p_{\text{s}}\big)\big(p^{2}+pq+q^{2}\big)},\hspace{1.4cm}
	\pi_{1,2}=\frac{p^{2}q\big(1-p_{\text{s}}\big)}{\big(p+q\big)\big(2-p_{\text{s}}\big)\big(p^2+pq+q^2\big)}\notag\\
	\pi_{2,0}&=\frac{p^{2}q\big(1-p_{\text{s}}\big)^{2}}{\big(p+q\big)\big(2-p_{\text{s}}\big)\big(p^2+pq+q^2\big)},\hspace{0.1cm}
	\pi_{2,1}=\frac{p^{2}\big(1-p_{\text{s}}\big)}{\big(2-p_{\text{s}}\big)\big(p^2+pq+q^2\big)}\notag\\
	\pi_{2,2}&=\frac{p^{2}\big[p+qp_{\text{s}}\big(2-p_{\text{s}}\big)\big]}{\big(p+q\big)\big(2-p_{\text{s}}\big)\big(p^{2}+pq+q^{2}\big)}.
\end{align}
For the semantics-aware policy $\pi_{i,j}$ can be calculated as
\begin{align}
	\label{pij_SA_M2_N3}
	\pi_{0,0}&=\frac{q^{2}\big[pp_{\text{s}}(1-p_{\text{s}})+\big(q+(1-q)p_{\text{s}}\big)\big]}{ \big(p^2+p q+q^2\big)\Big[p\big(1-p_{\text{s}}\big)\big(q+(2-q)p_{\text{s}}\big)+\big(q+p_{\text{s}}(1-q)\big)^{2}+p^{2}\big(1-p_{\text{s}}\big)^{2}\Big]}\notag\\
	\pi_{0,1} &=\pi_{1,0}= \frac{p q^{2}\big(1-p_{\text{s}}\big)\big[q+(1-q)p_{\text{s}}\big]}{ \big(p^2+p q+q^2\big)\Big[p\big(1-p_{\text{s}}\big)\big(q+(2-q)p_{\text{s}}\big)+\big(q+p_{\text{s}}(1-q)\big)^{2}+p^{2}\big(1-p_{\text{s}}\big)^{2}\Big]}\notag\\
	\pi_{1,1}&=\frac{pq\big[p_{\text{s}}+p(1-p_{\text{s}})\big]\big[q+(1-q)p_{\text{s}}\big]}{ \big(p^2+p q+q^2\big)\Big[p\big(1-p_{\text{s}}\big)\big(q+(2-q)p_{\text{s}}\big)+\big(q+p_{\text{s}}(1-q)\big)^{2}+p^{2}\big(1-p_{\text{s}}\big)^{2}\Big]}\notag\\
	\pi_{0,2}&=\pi_{2,0}=\frac{p^{2}q^{2}\big(1-p_{\text{s}}\big)^{2}}{ \big(p^2+p q+q^2\big)\Big[p\big(1-p_{\text{s}}\big)\big(q+(2-q)p_{\text{s}}\big)+\big(q+p_{\text{s}}(1-q)\big)^{2}+p^{2}\big(1-p_{\text{s}}\big)^{2}\Big]}\notag\\
	\pi_{1,2}&=\pi_{2,1}=\frac{p^{2}q\big(1-p_{\text{s}}\big)\big[p+(1-p)p_{\text{s}}\big]}{ \big(p^2+p q+q^2\big)\Big[p\big(1-p_{\text{s}}\big)\big(q+(2-q)p_{\text{s}}\big)+\big(q+p_{\text{s}}(1-q)\big)^{2}+p^{2}\big(1-p_{\text{s}}\big)^{2}\Big]}\notag\\
	\pi_{2,2}&=\frac{p^{2}\big[2pp_{\text{s}}(1-p_{\text{s}})+p^{2}(1-p_{\text{s}})^{2}+p_{\text{s}}\big(q+(1-q)p_{\text{s}}\big)\big]}{\big(p^2+p q+q^2\big)\Big[p\big(1-p_{\text{s}}\big)\big(q+(2-q)p_{\text{s}}\big)+\big(q+p_{\text{s}}(1-q)\big)^{2}+p^{2}\big(1-p_{\text{s}}\big)^{2}\Big]}.
\end{align}

\subsection{Proof of Lemma {\ref{Optimization1_lemma1}}} 
\label{Appendix_Optimization1_Lemma1}
To prove, we need to compare the objective function given in \eqref{PE_M1_N2_2} with the time-averaged reconstruction error when no sampling is performed, that is, $P^{\text{NS}}_{E} =1/2$. 

In this case, it can be readily shown that the objective function is smaller than $P^{\text{NS}}_{E}$ for all values of $p_{\text{s}}$ and $p_{\alpha^{\text{s}}}$. In addition, the objective function is decreasing with $p_{\alpha^{\text{s}}}$, i.e., $\frac{\partial P_{E}}{\partial p_{\alpha^{\text{s}}}}<0$ for all values of $p$ and $p_{\text{s}}$. In other words, the objective function has its minimum value when $p_{\alpha^{\text{s}}}$ is maximum. Now, using the constraint given in \eqref{Optimization_problem1_constraint}, the optimal value of sampling probability $p^{*}_{\alpha^{\text{s}}}$ is $\eta$. Therefore, the minimum value of the optimization problem, $P^{*}_{E}$, is obtained as
\begin{align}
	\label{PE_min1}
	P^{*}_{E} =  \frac{2\big(p-p  p_{\text{s}}\eta\big)}{4p+2p_{\text{s}}\eta-4pp_{\text{s}}\eta}.
\end{align}

\subsection{Proof of Lemma {\ref{Optimization2_lemma1}}}
\label{Appendix_Optimization2_Lemma1}
To address the optimization problem given in \eqref{OptimizationProb2}, we assume that $P^{\text{AS}}_{E}<P^{\text{NS}}_{E}$ because when $P^{\text{AS}}_{E}\geqslant P^{\text{NS}}_{E}$ the optimal policy is to not perform sampling. Now, to calculate $\bar{C}_{E}(n)$ we need to obtain the stationary distribution of the DTMC in Fig. \ref{Fig_ConsecutiveError2}. Based on this figure, for a fixed $n\in \mathbb{N}$, we can obtain the steady state probabilities of the consecutive error as follows
\begin{align}
	\label{ConsecutiveError_Pks}
	\pi_{k}(n) &=
	\begin{cases}
		\displaystyle \big(P^{\text{NS}}_{E}\big)^{k}\pi_{0}, \hspace{0.2cm} &1\leqslant k \leqslant n-1\\
		\displaystyle\big(P^{\text{AS}}_{E}\big)^{k-n}\big(P^{\text{NS}}_{E}\big)^{n}\pi_{0}, \hspace{0.2cm} & k\geqslant n,
	\end{cases} 
\end{align}
where
\begin{align}
	\label{ConsecutiveError_P0}
	\pi_{0}(n) = \Bigg[1+\frac{P^{\text{NS}}_{E}-\big(P^{\text{NS}}_{E}\big)^{n}}{1-P^{\text{NS}}_{E}}+\frac{\big(P^{\text{NS}}_{E}\big)^{n}}{1-P^{\text{AS}}_{E}}\Bigg]^{-1}.
\end{align}
Now, using \eqref{ConsecutiveError_Pks} and \eqref{ConsecutiveError_P0}, we can obtain the average $\bar{C}_{E}(n)$ as 
\begin{align}
	\label{CM_E}
	\bar{C}_{E}(n) &= \sum_{k=1}^{\infty}k\pi_{k}(n)\notag\\
	&=\left[\frac{P^{\text{NS}}_{E}-\big(P^{\text{NS}}_{E}\big)^{n+1}-n\big(P^{\text{NS}}_{E}\big)^{n}+n\big(P^{\text{NS}}_{E}\big)^{n+1}}{\Big(1-P^{\text{NS}}_{E}\Big)^{2}}+\frac{\big(P^{\text{NS}}_{E}\big)^{n}\Big(n+P^{\text{AS}}_{E}-n P^{\text{AS}}_{E}\Big)}{\Big(1-P^{\text{AS}}_{E}\Big)^{2}}\right]\notag\\
	&\times \Bigg[1+\frac{P^{\text{NS}}_{E}-\big(P^{\text{NS}}_{E}\big)^{n}}{1-P^{\text{NS}}_{E}}+\frac{\big(P^{\text{NS}}_{E}\big)^{n}}{1-P^{\text{AS}}_{E}}\Bigg]^{-1}.
\end{align}
Note that the constraint given in \eqref{Optimization_problem2_constraint} can be described as a portion of time where the sampler always performs sampling. Hence, we can write
\begin{align}
	\label{Tn}
	\frac{1}{T}\sum_{t=1}^{T}\delta \mathbbm{1}\{\alpha^{\text{s}}_{t}=1\} = \sum_{k=n}^{\infty}\delta\pi_{k}(n) = \frac{\delta\big(P^{\text{NS}}_{E}\big)^{n}}{1-P^{\text{AS}}_{E}}\times \Bigg[1+\frac{P^{\text{NS}}_{E}-\big(P^{\text{NS}}_{E}\big)^{n}}{1-P^{\text{NS}}_{E}}+\frac{\big(P^{\text{NS}}_{E}\big)^{n}}{1-P^{\text{AS}}_{E}}\Bigg]^{-1}.
\end{align}
Since we assume that $P^{\text{AS}}_{E}<P^{\text{NS}}_{E}$, $\bar{C}_{E}(n)$ in \eqref{CM_E} is increasing with $n$. Therefore, the optimal value of $n$ that minimizes $\bar{C}_{E}(n)$ is the minimum value of $n$ that satisfies the constraint given in \eqref{Optimization_problem2_constraint}. Now, using \eqref{Tn}, we can write the constraint as
\begin{align}
	\label{Prob2_Constraint}
	\frac{\big(P^{\text{NS}}_{E}\big)^{n}}{1-P^{\text{AS}}_{E}}\times \Bigg[1+\frac{P^{\text{NS}}_{E}-\big(P^{\text{NS}}_{E}\big)^{n}}{1-P^{\text{NS}}_{E}}+\frac{\big(P^{\text{NS}}_{E}\big)^{n}}{1-P^{\text{AS}}_{E}}\Bigg]^{-1}\leqslant \eta,
\end{align}
where $\eta = \frac{\delta_{\text{max}}}{\delta}$. Simplifying inequality \eqref{Prob2_Constraint} we obtain the optimal value of $n$ that minimizes $\bar{C}^{\text{M}}_{E}(n)$ as
\begin{align}
	\label{n_optimal_1}
	n^{*} = \left\lceil{\log_{P^{\text{NS}}_{E}}\left(\frac{\eta\big(1-P^{\text{AS}}_{E}\big)}{1-(1-\eta)P^{\text{NS}}_{E}-\eta P^{\text{AS}}_{E}}\right)}\right\rceil.
\end{align}

\subsection{Time-averaged sampling cost for the semantics-aware and change-aware policies}
\label{Appendix_AvgCost}
\par The time-averaged sampling cost in \eqref{Optimization_prob1_constraint} for the semantics-aware policy can be written as 
\begin{align}
    \label{semantic_samplingcost_M1}
    \lim_{T \to \infty}\frac{1}{T}\sum_{t=1}^{T}\delta \mathbbm{1}\{\alpha^{\text{s}}_{t}=1\} &= \delta\mathrm{Pr}\big[X_{t+1}\neq X_{t}\big|E_{t} = 0\big]\mathrm{Pr}\big[E_{t} = 0\big]\notag\\&+\delta\mathrm{Pr}\big[X_{t+1}\neq \hat{X}_{t}\big|E_{t} \neq0\big]\mathrm{Pr}\big[E_{t} \neq 0\big],
\end{align}
where using two-state DTMC model, \eqref{semantic_samplingcost_M1} can be written as
\begin{align}
    \label{semantic_samplingcost_M1_2}
    \lim_{T \to \infty}\frac{1}{T}\sum_{t=1}^{T}\delta \mathbbm{1}\{\alpha^{\text{s}}_{t}=1\} &=\delta\mathrm{Pr}\big[X_{t+1}=1, X_{t} = 0, \hat{X}_{t} =0\big]+\delta\mathrm{Pr}\big[X_{t+1}=0, X_{t} = 1, \hat{X}_{t} =1\big]\notag\\
    &+\delta\mathrm{Pr}\big[X_{t+1}=0, X_{t} = 0, \hat{X}_{t} =1\big]+\delta\mathrm{Pr}\big[X_{t+1}=1, X_{t} = 1, \hat{X}_{t} =0\big]\notag\\
    &=\delta p\big(\pi_{0,0}+\pi_{1,1}\big)+\delta(1-p)\big(\pi_{0,1}+\pi_{1,0}\big) = \frac{2p\delta}{4p+2p_{\text{s}}-4pp_{\text{s}}}.
\end{align}
For the change-aware policy, we can write \eqref{semantic_samplingcost_M1} as
\begin{align}
    \label{ChangeAware_samplingcost_M1}
    \lim_{T \to \infty}\frac{1}{T}\sum_{t=1}^{T}\delta \mathbbm{1}\{\alpha^{\text{s}}_{t}=1\} &= \delta\mathrm{Pr}\big[X_{t+1}\neq X_{t}\big|E_{t} = 0\big]\mathrm{Pr}\big[E_{t} = 0\big]\notag\\&+\delta\mathrm{Pr}\big[X_{t+1}\neq {X}_{t}\big|E_{t} \neq0\big]\mathrm{Pr}\big[E_{t} \neq 0\big],
\end{align}
where for two-state DTMC, \eqref{ChangeAware_samplingcost_M1} is calculated as
\begin{align}
    \label{ChangeAware_samplingcost_M1_2}
    \lim_{T \to \infty}\frac{1}{T}\sum_{t=1}^{T}\delta \mathbbm{1}\{\alpha^{\text{s}}_{t}=1\} &=\delta\mathrm{Pr}\big[X_{t+1}=1, X_{t} = 0, \hat{X}_{t} =0\big]+\delta\mathrm{Pr}\big[X_{t+1}=0, X_{t} = 1, \hat{X}_{t} =1\big]\notag\\
    &+\delta\mathrm{Pr}\big[X_{t+1}=1, X_{t} = 0, \hat{X}_{t} =1\big]+\delta\mathrm{Pr}\big[X_{t+1}=0, X_{t} = 1, \hat{X}_{t} =0\big]\notag\\
    &=\delta p.
\end{align}
Considering now the BDMP model and using a similar procedure presented in \eqref{semantic_samplingcost_M1}, the time-averaged sampling cost given in \eqref{Optimization_prob1_constraint} for the semantics-aware policy can be written as
\begin{align}
    \label{semantic_samplingcost_M2}
    \lim_{T \to \infty}\frac{1}{T}\sum_{t=1}^{T}\delta \mathbbm{1}\{\alpha^{\text{s}}_{t}=1\}&=\delta p \pi_{0,0}+\delta q\pi_{1,1}+\delta (1-p)\pi_{0,1}+\delta(1-q)\pi_{1,0}\notag\\ &= \frac{2pq\delta}{(p+q)\big[p(1-p_{\text{s}})+q+p_{\text{s}}(1-q)\big]}.
\end{align}
Furthermore, for the change-aware policy we can write 
\begin{align}
    \label{ChangeAware_samplingcost_M2}
    \lim_{T \to \infty}\frac{1}{T}\sum_{t=1}^{T}\delta \mathbbm{1}\{\alpha^{\text{s}}_{t}=1\} &= \delta p\big(\pi_{0,0}+\pi_{0,1}\big)+\delta q\big(\pi_{1,1}+\pi_{1,0}\big) = \frac{2pq\delta}{p+q}.
\end{align}
\bibliographystyle{IEEEtran}
\bibliography{ref}

\begin{thebibliography}{10}
\providecommand{\url}[1]{#1}
\csname url@samestyle\endcsname
\providecommand{\newblock}{\relax}
\providecommand{\bibinfo}[2]{#2}
\providecommand{\BIBentrySTDinterwordspacing}{\spaceskip=0pt\relax}
\providecommand{\BIBentryALTinterwordstretchfactor}{4}
\providecommand{\BIBentryALTinterwordspacing}{\spaceskip=\fontdimen2\font plus
\BIBentryALTinterwordstretchfactor\fontdimen3\font minus
  \fontdimen4\font\relax}
\providecommand{\BIBforeignlanguage}[2]{{%
\expandafter\ifx\csname l@#1\endcsname\relax
\typeout{** WARNING: IEEEtran.bst: No hyphenation pattern has been}%
\typeout{** loaded for the language `#1'. Using the pattern for}%
\typeout{** the default language instead.}%
\else
\language=\csname l@#1\endcsname
\fi
#2}}
\providecommand{\BIBdecl}{\relax}
\BIBdecl

\bibitem{xu2004optimal}
Y.~Xu and J.~P. Hespanha, ``{Optimal communication logics in networked control
  systems},'' in \emph{IEEE Conference on Decision and Control (CDC)}, 2004.

\bibitem{shi2011sensor}
L.~Shi, P.~Cheng, and J.~Chen, ``{Sensor data scheduling for optimal state
  estimation with communication energy constraint},'' \emph{Automatica},
  vol.~47, no.~8, pp. 1693--1698, 2011.

\bibitem{shi2011time}
L.~Shi, K.~H. Johansson, and L.~Qiu, ``{Time and event-based sensor scheduling
  for networks with limited communication resources},'' \emph{IFAC Proceedings
  Volumes}, vol.~44, no.~1, pp. 13\,263--13\,268, 2011.

\bibitem{wu2013can}
J.~Wu, Y.~Yuan, H.~Zhang, and L.~Shi, ``{How can online schedules improve
  communication and estimation tradeoff?}'' \emph{IEEE Transactions on Signal
  Processing}, vol.~61, no.~7, pp. 1625--1631, 2013.

\bibitem{li2010event}
L.~Li, M.~Lemmon, and X.~Wang, ``{Event-triggered state estimation in vector
  linear processes},'' in \emph{IEEE American Control Conference (ACC)}, 2010.

\bibitem{VictorNTNU}
V.~W. Håkansson, N.~K.~D. Venkategowda, and S.~Werner, ``{Optimal Transmission
  Threshold and Channel Allocation Strategies for Heterogeneous Sensor Data},''
  in \emph{55th Asilomar Conference on Signals, Systems, and Computers}, 2021.

\bibitem{wu2012event}
J.~Wu, Q.-S. Jia, K.~H. Johansson, and L.~Shi, ``{Event-based sensor data
  scheduling: Trade-off between communication rate and estimation quality},''
  \emph{IEEE Transactions on Automatic Control}, vol.~58, no.~4, pp.
  1041--1046, 2012.

\bibitem{weimer2012distributed}
J.~Weimer, J.~Ara{\'u}jo, and K.~H. Johansson, ``{Distributed event-triggered
  estimation in networked systems},'' \emph{IFAC Proceedings Volumes}, vol.~45,
  no.~9, pp. 178--185, 2012.

\bibitem{meng2012optimal}
X.~Meng and T.~Chen, ``{Optimal sampling and performance comparison of periodic
  and event based impulse control},'' \emph{IEEE Transactions on Automatic
  Control}, vol.~57, no.~12, pp. 3252--3259, 2012.

\bibitem{trimpe2014event}
S.~Trimpe and R.~D'Andrea, ``{Event-based state estimation with variance-based
  triggering},'' \emph{IEEE Transactions on Automatic Control}, vol.~59,
  no.~12, 2014.

\bibitem{leong2016sensor}
A.~S. Leong, S.~Dey, and D.~E. Quevedo, ``{Sensor scheduling in variance based
  event triggered estimation with packet drops},'' \emph{IEEE Transactions on
  Automatic Control}, vol.~62, no.~4, pp. 1880--1895, 2016.

\bibitem{imer2005optimal}
O.~C. Imer and T.~Basar, ``{Optimal estimation with limited measurements},'' in
  \emph{IEEE Conference on Decision and Control (CDC)}, 2005.

\bibitem{nayyar2013optimal}
A.~Nayyar, T.~Ba{\c{s}}ar, D.~Teneketzis, and V.~V. Veeravalli, ``{Optimal
  strategies for communication and remote estimation with an energy harvesting
  sensor},'' \emph{IEEE Transactions on Automatic Control}, vol.~58, no.~9, pp.
  2246--2260, 2013.

\bibitem{chakravorty2014optimal}
J.~Chakravorty and A.~Mahajan, ``{On the optimal thresholds in remote state
  estimation with communication costs},'' in \emph{IEEE Conference on Decision
  and Control (CDC)}, 2014.

\bibitem{shi2012scheduling}
L.~Shi and H.~Zhang, ``{Scheduling two {Gauss--Markov} systems: An optimal
  solution for remote state estimation under bandwidth constraint},''
  \emph{IEEE Transactions on Signal Processing}, vol.~60, no.~4, pp.
  2038--2042, 2012.

\bibitem{wu2018optimal}
S.~Wu, X.~Ren, S.~Dey, and L.~Shi, ``{Optimal scheduling of multiple sensors
  over shared channels with packet transmission constraint},''
  \emph{Automatica}, vol.~96, pp. 22--31, 2018.

\bibitem{chakravorty2015distortion}
J.~Chakravorty and A.~Mahajan, ``{Distortion-transmission trade-off in
  real-time transmission of Markov sources},'' in \emph{IEEE Information Theory
  Workshop (ITW)}, 2015.

\bibitem{chakravorty2016fundamental}
------, ``{Fundamental limits of remote estimation of autoregressive Markov
  processes under communication constraints},'' \emph{IEEE Transactions on
  Automatic Control}, vol.~62, no.~3, pp. 1109--1124, 2016.

\bibitem{sun2019sampling}
Y.~Sun, Y.~Polyanskiy, and E.~Uysal, ``{Sampling of the Wiener process for
  remote estimation over a channel with random delay},'' \emph{IEEE
  Transactions on Information Theory}, vol.~66, no.~2, pp. 1118--1135, 2019.

\bibitem{ornee2021sampling}
T.~Z. Ornee and Y.~Sun, ``{Sampling and remote estimation for the
  ornstein-uhlenbeck process through queues: Age of information and beyond},''
  \emph{IEEE/ACM Transactions on Networking}, vol.~29, no.~5, pp. 1962--1975,
  2021.

\bibitem{GuoKostina2022}
N.~Guo and V.~Kostina, ``{Optimal Causal Rate-Constrained Sampling for a Class
  of Continuous Markov Processes},'' \emph{IEEE Transactions on Information
  Theory}, vol.~67, no.~12, pp. 7876--7890, 2021.

\bibitem{hui2022real}
H.~Hui, S.~Hu, and W.~Chen, ``{Real Time Monitoring of Brownian Motions},''
  \emph{IEEE Transactions on Communications}, vol.~70, no.~9, pp. 5867--5881,
  2022.

\bibitem{chakravorty2019remote}
J.~Chakravorty and A.~Mahajan, ``{Remote estimation over a packet-drop channel
  with Markovian state},'' \emph{IEEE Transactions on Automatic Control},
  vol.~65, no.~5, pp. 2016--2031, 2019.

\bibitem{lipsa2011remote}
G.~M. Lipsa and N.~C. Martins, ``{Remote state estimation with communication
  costs for first-order LTI systems},'' \emph{IEEE Transactions on Automatic
  Control}, vol.~56, no.~9, pp. 2013--2025, 2011.

\bibitem{huang2019retransmit}
K.~Huang, W.~Liu, Y.~Li, and B.~Vucetic, ``{To retransmit or not: Real-time
  remote estimation in wireless networked control},'' in \emph{IEEE
  International Conference on Communications (ICC)}, 2019.

\bibitem{pezzutto2022transmission}
M.~Pezzutto, L.~Schenato, and S.~Dey, ``{Transmission power allocation for
  remote estimation with multi-packet reception capabilities},''
  \emph{Automatica}, vol. 140, p. 110257, 2022.

\bibitem{ShannonWeaver49}
C.~E. Shannon and W.~Weaver, \emph{{The mathematical theory of
  communication}}.\hskip 1em plus 0.5em minus 0.4em\relax University of
  Illinois Press, Urbana, 1949.

\bibitem{VoI_USSR}
R.~L. {Stratonovich}, ``{On the value of information},'' \emph{Izv. USSR Acad.
  Sci. Tech. Cybern.}, no.~5, 1965.

\bibitem{QoI}
C.~Bisdikian, L.~M. Kaplan, and M.~B. Srivastava, ``{On the Quality and Value
  of Information in Sensor Networks},'' \emph{ACM Transactions on Sensor
  Networks}, vol.~9, no.~4, July 2013.

\bibitem{kountouris2021semantics}
M.~Kountouris and N.~Pappas, ``{Semantics-empowered communication for networked
  intelligent systems},'' \emph{IEEE Communications Magazine}, vol.~59, no.~6,
  pp. 96--102, 2021.

\bibitem{Qin22arxiv}
Z.~Qin, X.~Tao, J.~Lu, and G.~Y. Li, ``{Semantic Communications: Principles and
  Challenges},'' \emph{arXiv preprint arXiv: 2201.01389v2}, 2022.

\bibitem{tolga21SP}
M.~Kalfa, M.~Gok, A.~Atalik, B.~Tegin, T.~M. Duman, and O.~Arikan, ``{Towards
  goal-oriented semantic signal processing: Applications and future
  challenges},'' \emph{Digital Signal Processing}, vol. 119, 2021.

\bibitem{PetarProc2022}
P.~Popovski and \textit{et al.}, ``{A Perspective on Time Toward Wireless
  6{G}},'' \emph{Proceedings of the IEEE}, vol. 110, no.~8, 2022.

\bibitem{GunduzJSAC23}
D.~Gündüz, Z.~Qin, I.~E. Aguerri, H.~S. Dhillon, Z.~Yang, A.~Yener, K.~K.
  Wong, and C.-B. Chae, ``{Beyond Transmitting Bits: Context, Semantics, and
  Task-Oriented Communications},'' \emph{IEEE J. Sel. Areas Commun.}, vol.~41,
  no.~1, pp. 5--41, 2023.

\bibitem{pappas2021goal}
N.~Pappas and M.~Kountouris, ``{Goal-oriented communication for real-time
  tracking in autonomous systems},'' in \emph{IEEE International Conference on
  Autonomous Systems (ICAS)}, 2021.

\end{thebibliography}
\end{document}